\documentclass{article}

\usepackage{graphicx}
\usepackage{subcaption}
\usepackage{authblk}

\title{Multifunctional synaptic learning and neuromorphic computing using crystalline TiO$_x$/NiO$_x$ heterojunction-based memory devices}

\author[1,2]{Dip Manna}
\author[1,2]{Koushik Mondal}
\author[3]{Subhajit Karmakar}
\author[1]{Pawan Kumar Ojha}
\author[4,2]{Sourav Roy}
\author[1]{Biswarup Satpati}
\author[4]{Prasanta Karmakar}
\author[1]{Supratic Chakraborty}

\affil[1]{Saha Institute of Nuclear Physics, Kolkata, West Bengal, India}
\affil[2]{Homi Bhabha National Institute, Mumbai, Maharashtra, India}
\affil[3]{University Science Instrumentation Centre, The University of Burdwan, Burdwan, West Bengal, India}
\affil[4]{Variable Energy Cyclotron Centre, Kolkata, West Bengal, India}

\date{}

\begin{document}

\maketitle

\maketitle

\vspace{10pt}


\abstract {An attempt is made here to mimic different properties of biological synapses using Ag/TiO$_x$/NiO$_x$/p$^{++}$-Si memristor structure by studying its different transport properties under dc and pulsed bias. The presence of crystalline NiO$_x$ with smaller gains is found to be helpful to get TiO$_x$ deposited on its top with crystalline properties and larger grains. The heterostructure offers stable bipolar forming free non-volatile resistive switching characteristics under reverse biased condition with gradual set and reset features. These devices are also able to successfully implement the classical Pavlov's learning, study artificial nociceptor and Morse code detection. While incorporating synaptic weights derived from the measured conductance programming pulse relationship, an artificial neural network achieves nearly 95\% accuracy in MNIST digit recognition. In brief, NiO$_x$/TiO$_x$ heterojunction plays the pivotal role in getting such reproducible $I-V$ characteristics for mimicking different properties of biological synapses.}

\vspace{5pt}
\section{Introduction}
\label{sec1}
The human brain along with the spinal cord comprises the central nervous system where the brain is considered to be its central organ. Neurons are the fundamental units of the brain and nervous system, responsible for receiving sensory input from the external world. Neurons also send various signals to the muscles by transforming and relaying electrical signals. A pair of neurons is connected by a functional junction called synapse\cite{Guo}. The strength of the connection between two neurons, also termed as Synaptic weight plays an important role for information storage and its processing by its proper tuning\cite{Pereda}. Although the speed of a neuron or synapse is much slower than that of a transistor, parallel information processing in neural networks of a human body makes the human brain highly efficient compared to von Neumann computers for complex and unstructured information processing \cite{He,Yang}. Such an exceptional quality of the human brain encourages researchers to mimic the performance of the brain by developing similar computing hardware \cite{Silver}.  Fabrication of artificial synaptic devices have gained its pace for simulation of synaptic functions and become very popular for their different superior properties \cite{Upadhyay, Zhao}.  In analogy to the human brains, the neuromorphic computing system consists of artificial synapses acting as the fundamental element. The synapses store input information prompted by stimuli in the form of single/successive electric signal in the active layers of the synaptic devices and subsequently process the signal through  continuous updating of synaptic weights. Devices have so far been proposed based on different operating principles for mimcking the synaptic properties namely, resistive random access memory (ReRAM), phase change devices (PCM), transistors etc. \cite{Zhou}. Recently, two terminal metal-insulator-metal (MIM) architecture for emulating artificial synapses in the form of vertical memristor structure for neuromorphic computing has been an acceptable structure among the researchers due to its various advantages in the form of lesser power consumption, structurally simple, smaller size and ease of integration \cite{Fuller}. Their capability to modulate and retain multiple resistance states based on resistive switching (RS) principle provides an effective approach for emulating the adaptive behaviour of biological synapses. RS refers to the reversible modulation of the electrical resistance of a device between distinct resistance states upon the application of an external voltage. In many memristive systems, this behaviour originates from the electrically driven redistribution of defects, ions or charge carriers within the active layer led to formation or disruption of localized conductive paths. The resulting controllable changes in conductance can mimic key characteristics of biological synapses making RS devices attractive for hardware-based neuromorphic computing and artificial neural networks (ANN). By appropriately controlling their conductance states, these devices can emulate a variety of synaptic functions namely, long term potentiation (LTP) and depression (LTD), short term and long term memory (STM/LTM), spike timing dependent plasticity (STDP) and paired pulse facilitation (PPF). Such capabilities make memristors particularly suitable for implementing bioinspired learning and information processing functions in neuromorphic systems \cite{Li, Betal, Banerjee}. Along with these synaptic functions, the memristive device properties can also be used to mimic sensory functions of the nervous system nociception in particular that involves with the detection and processing of potentially harmful external stimuli through specialized biological receptors. These receptors exhibit characteristic responses such as threshold activation, stimulus dependent sensitivity and relaxation after the removal of the applied stimulus. The nonlinear and history dependent electrical response of memristive devices provides a basis for reproducing such behaviours in artificial nociceptive systems. Such bio-inspired nociceptive functionality can enable memristive devices to serve as sensory elements in neuromorphic systems bridging stimulus detection with subsequent information processing \cite{Dongyeo, Kumar}. 

In recent years, significant research efforts have focused on developing two terminal memristive devices based on diverse material platforms. Inorganic oxides, organic materials and composite structures have demonstrated distinct electrical and switching characteristics that can be tailored for different applications. This material diversity provides considerable scope for engineering memristive devices with their functional properties required for low power neuromorphic computing and bioinspired information processing \cite{Dutta, Dastgeer}. However, the organic and hybrid material-based memristive devices suffer from their limited thermal robustness and susceptibility to environmental conditions. These factors may lead to degradation of the active material and instability in the electrical response thereby affecting the long term reliability and operational lifetime of the devices \cite{wook, Mbumba}. In contrast, oxide-based materials generally offer superior thermal robustness and resistance to moisture induced degradation. Therefore, an appropriate selection and engineering of oxide materials can improve the environmental stability and operational reliability of memristive devices making them attractive candidates for robust neuromorphic applications \cite{Carlos}. Recent researches have highlighted the potential of metal oxide-based resistive switching memristors for neuromorphic applications. Their electrically tunable conductance, relatively low power consumption and compatibility with conventional CMOS fabrication technologies make them attractive candidates for implementing hardware-based synaptic functions and energy efficient neural information processing \cite{Khan, Panchanan, Tian, Ghazanfar}. Although single layer oxide memristors offer a relatively simple device architecture, their switching characteristics may be strongly influenced by the random formation and rupture of conductive filaments. This stochastic process may result in cycle to cycle and device to device variations making precise control of intermediate conductance states challenging \cite{Rabee, Ti, Pan}. To address these limitations,  bilayer and multilayer oxide architectures provide an additional degree of control over the switching process through formation of interfaces between two different oxide layers. Such interfacial engineering can regulate defect migration and conductive path evolution thereby improving switching stability, retention characteristics and the controllability of resistance states \cite{Wan, Xuerong, Hammad}. Similar to interface and barrier engineering approaches employed in conventional MIM-based resistive switching devices, the interface formed between $p$-type and $n$-type oxide layers can also offer an additional pathway for controlling resistive switching in metal-insulator-insulator-metal (MIIM) structures. The intrinsic rectifying behaviour of a $p-n$ heterojunction can establish asymmetric transport characteristics with relatively high and low resistance states under reverse and forward bias, respectively \cite{Zhen, Xiang}. Under suitable electrical stimulation, a controlled soft breakdown of the junction under reverse bias can induce a transition from the high resistance state (HRS) to the low resistance state (LRS). Conversely, applying a forward bias facilitates the recovery of the junction toward the HRS enabling reversible resistance modulation. Such interfacial control on carrier transport and defect migration provides an attractive strategy for achieving stable and controllable resistive switching in oxide heterostructures.

The TiO$_x$ and NiO$_x$ films were deposited by DC reactive sputtering followed by patterning of  top Ag electrodes. The resulting device exhibits gradual and reproducible resistance modulation with stable switching characteristics maintaining its functional behaviour for up to $10^{4}$~s over a period of 60 days. The results are presented and discussed here. 

\section{Results and Discussion}
\label{sec2}
\subsection{Structural, Morphological and Chemical Characterization}
\begin{figure*}[t]
\centering

\includegraphics[width=0.35\textwidth]{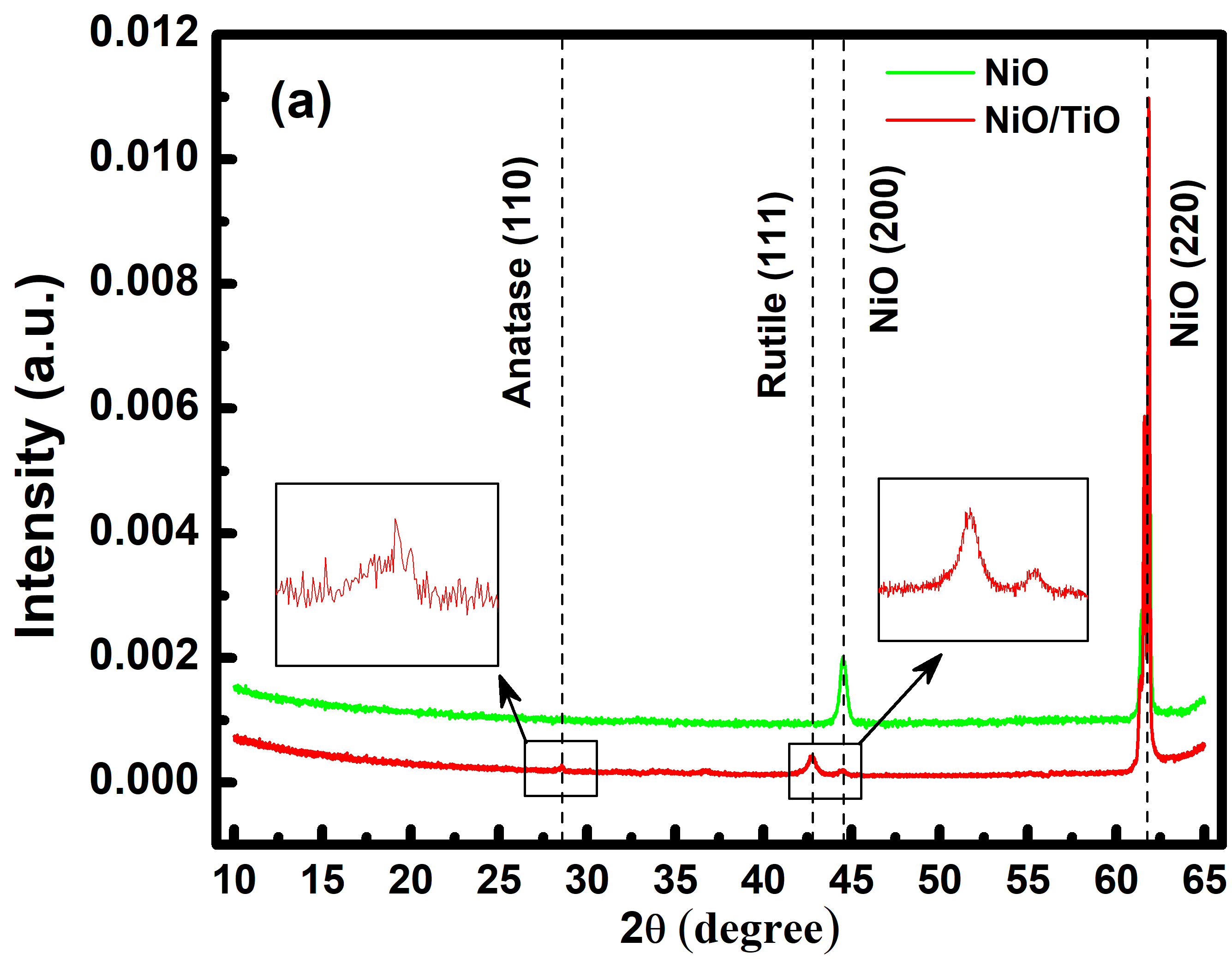}\hfill
\includegraphics[width=0.3\textwidth]{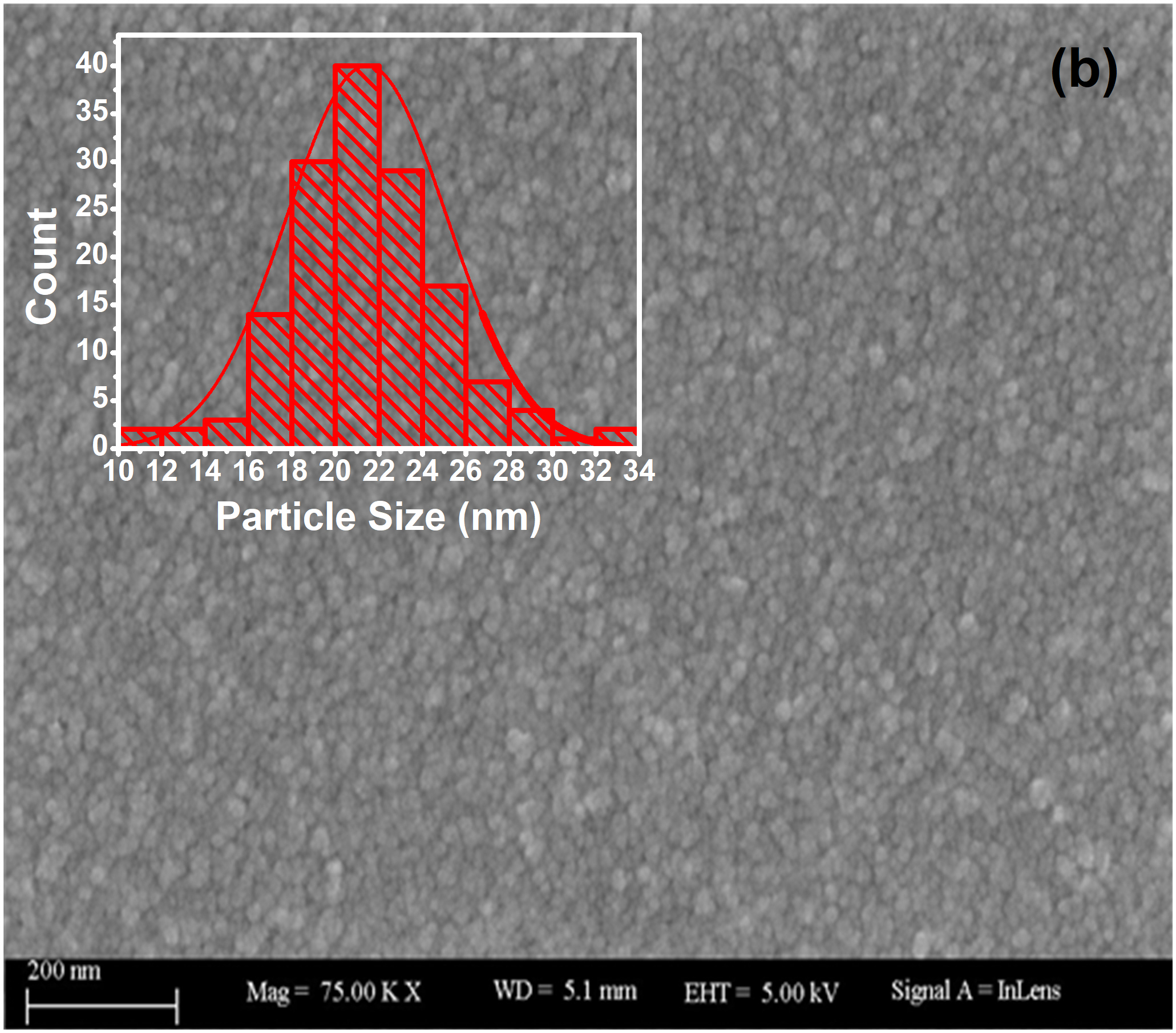}\hfill
\includegraphics[width=0.32\textwidth]{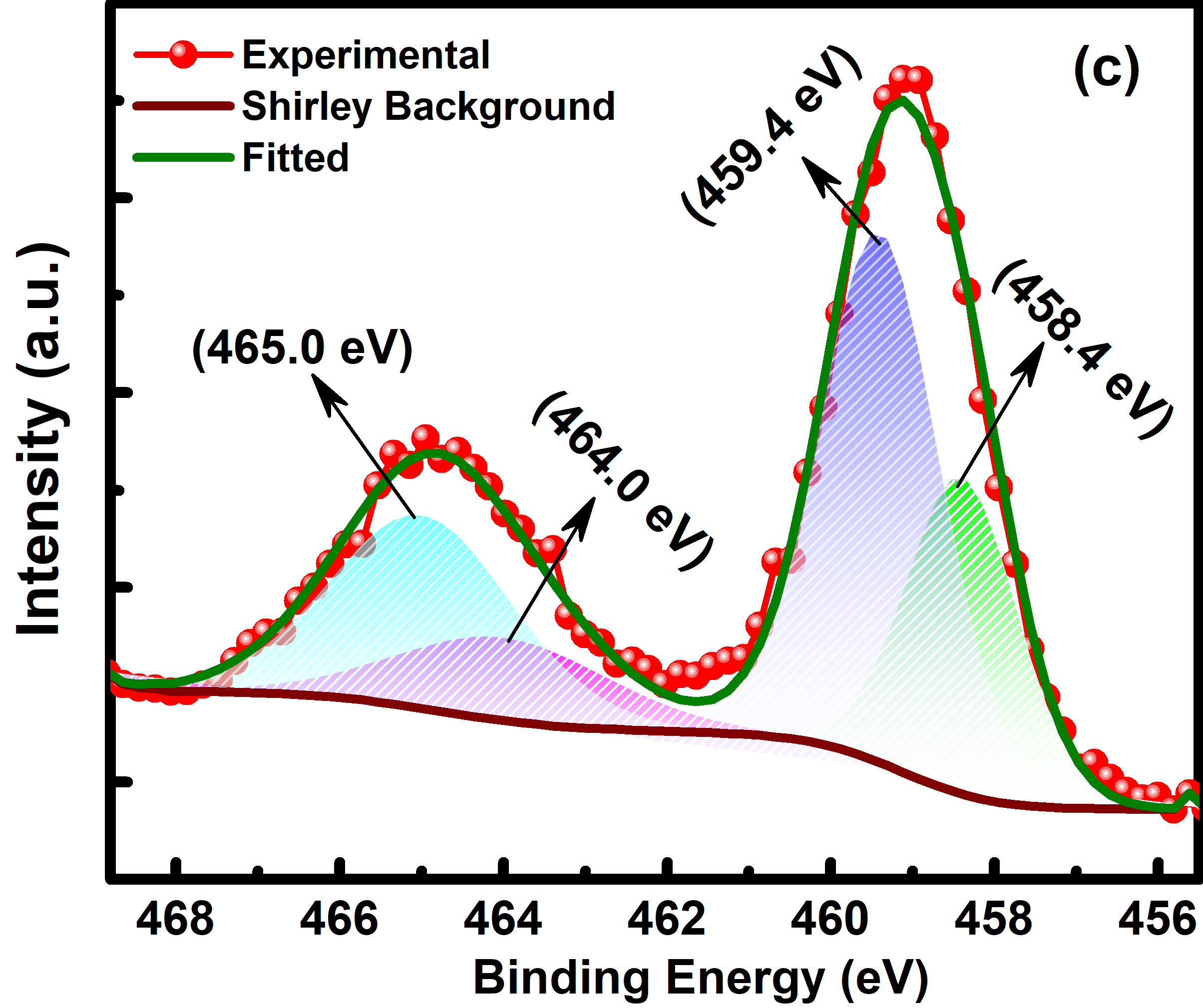}

\vspace{5mm}

\includegraphics[width=0.32\textwidth]{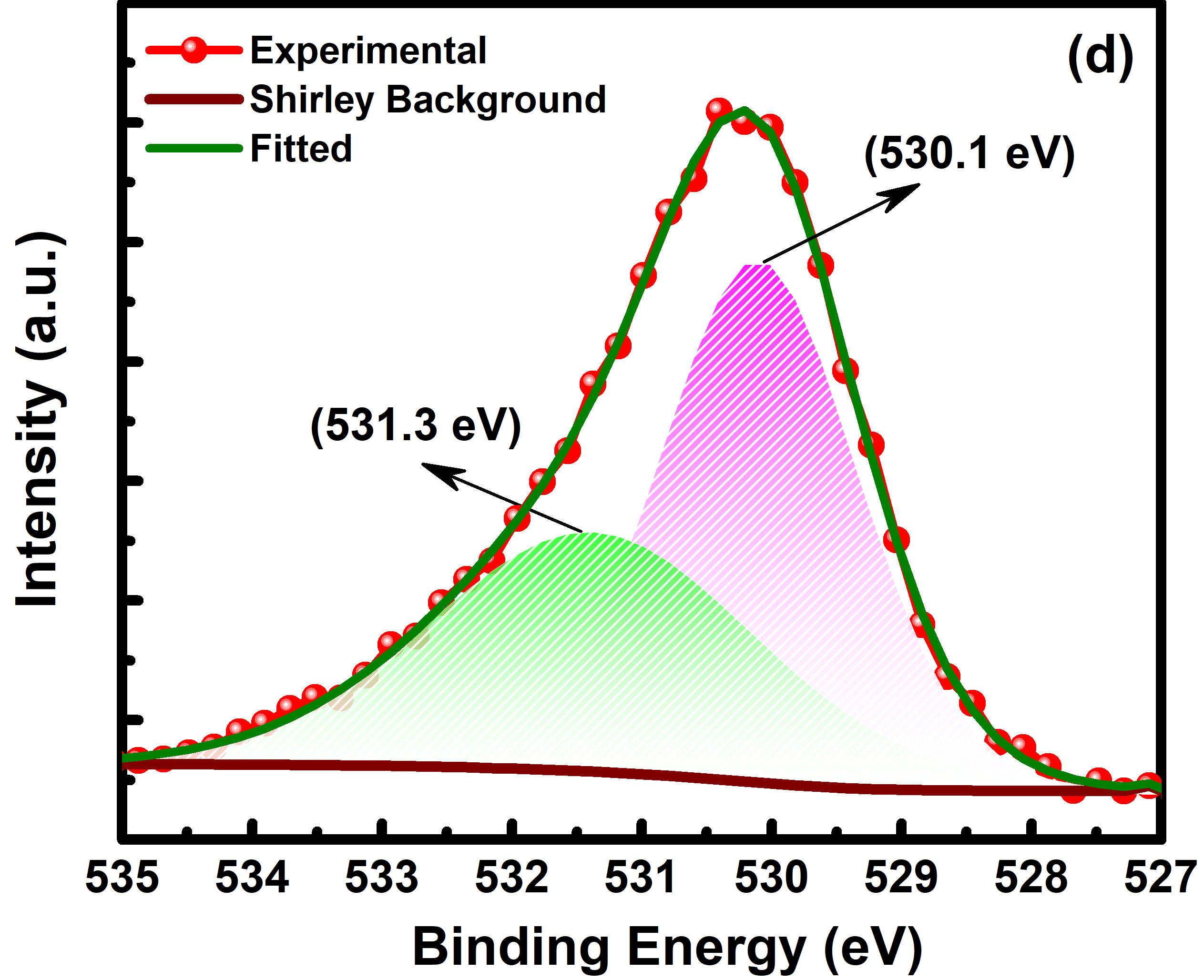}\hfill
\includegraphics[width=0.32\textwidth]{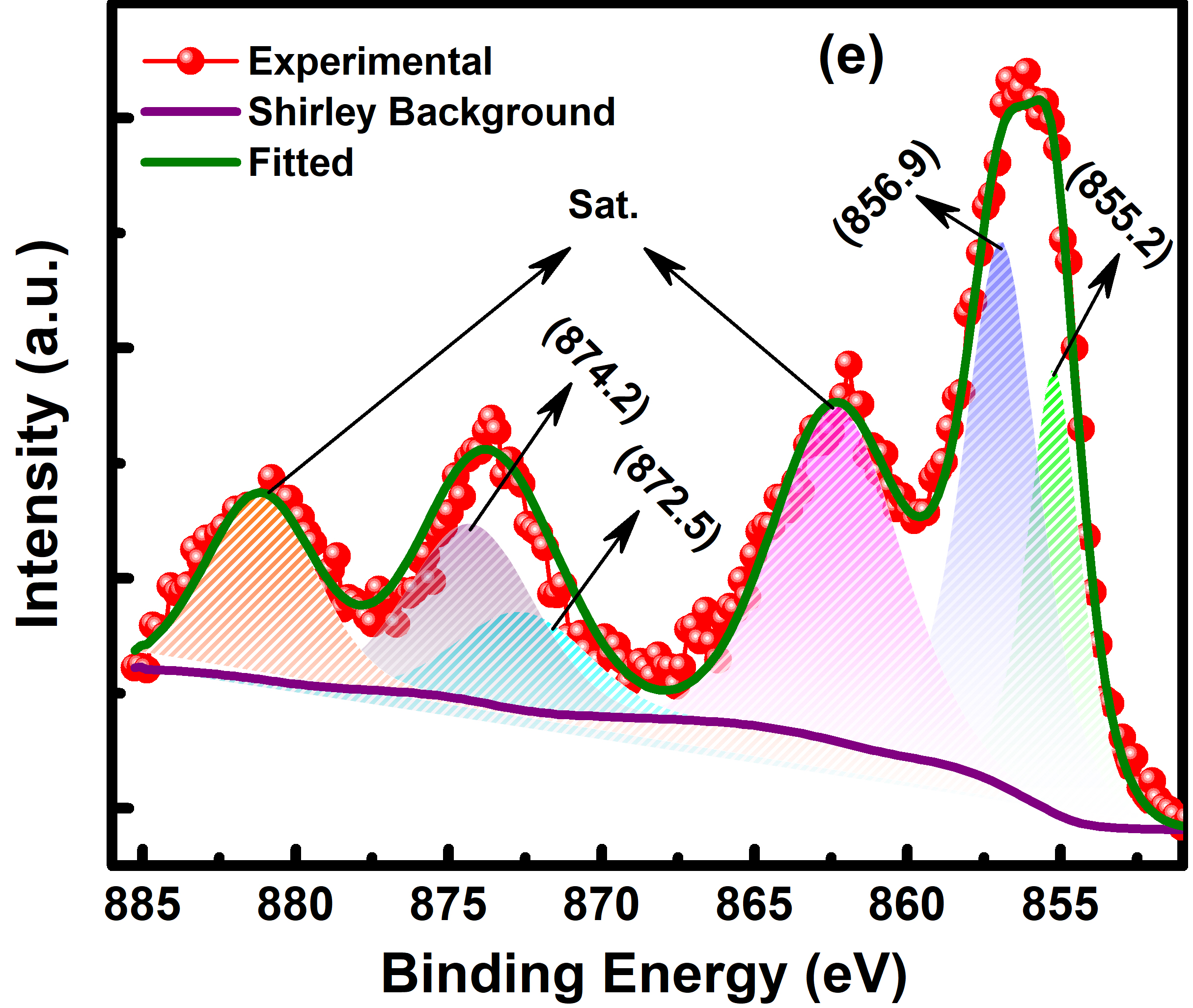}\hfill
\includegraphics[width=0.32\textwidth]{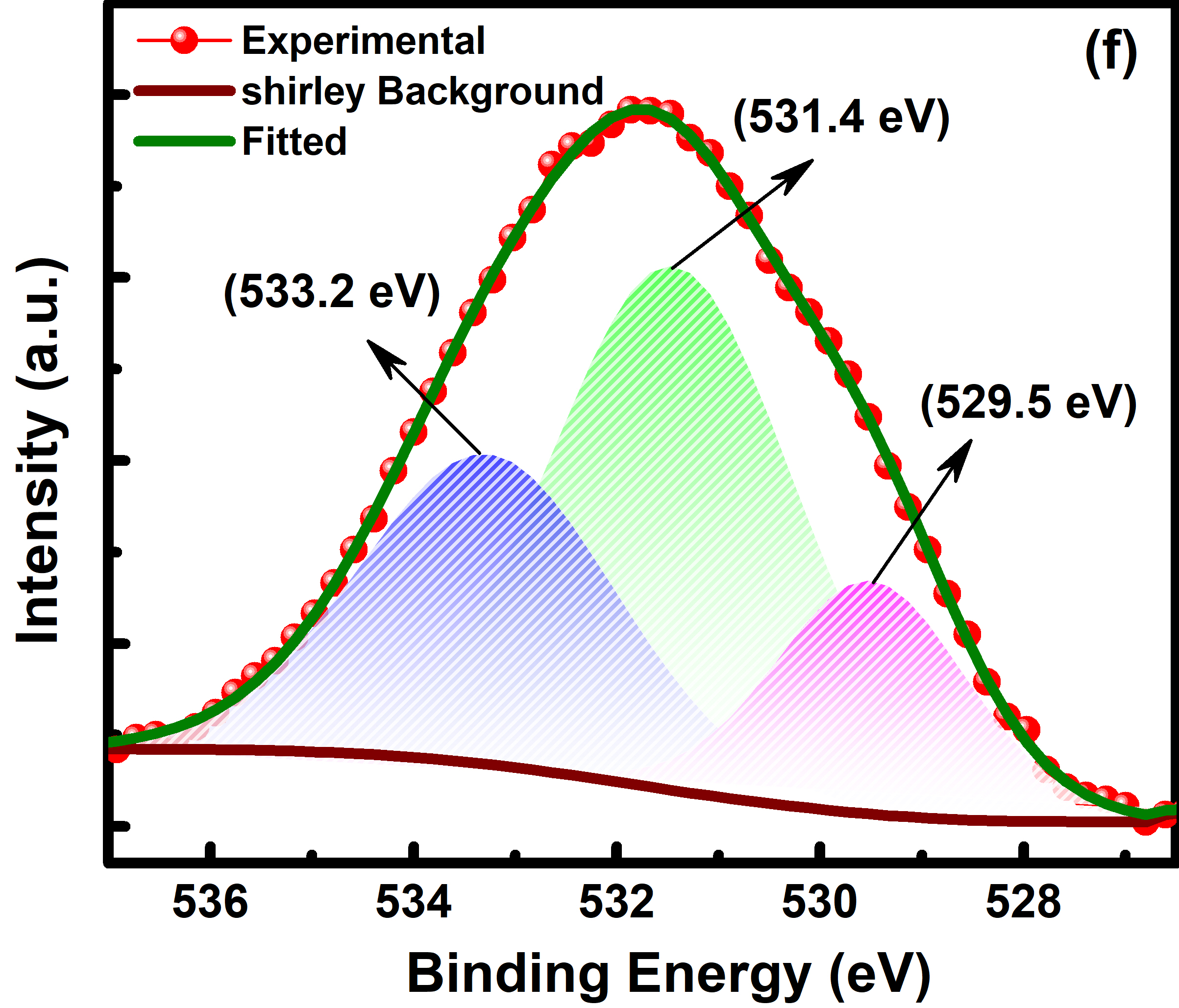}

\vspace{5mm}

\includegraphics[width=0.32\textwidth]{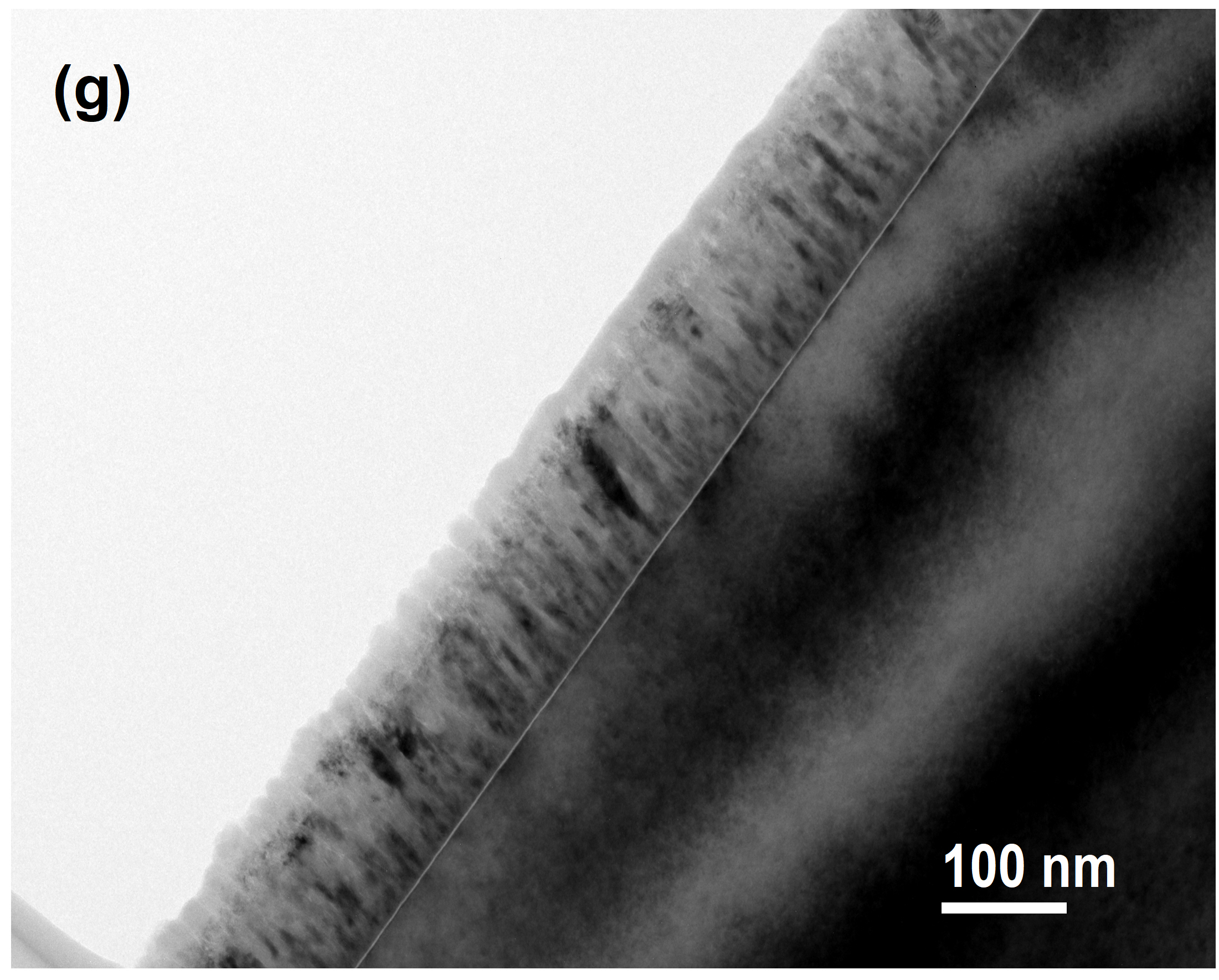}\hfill
\includegraphics[width=0.315\textwidth]{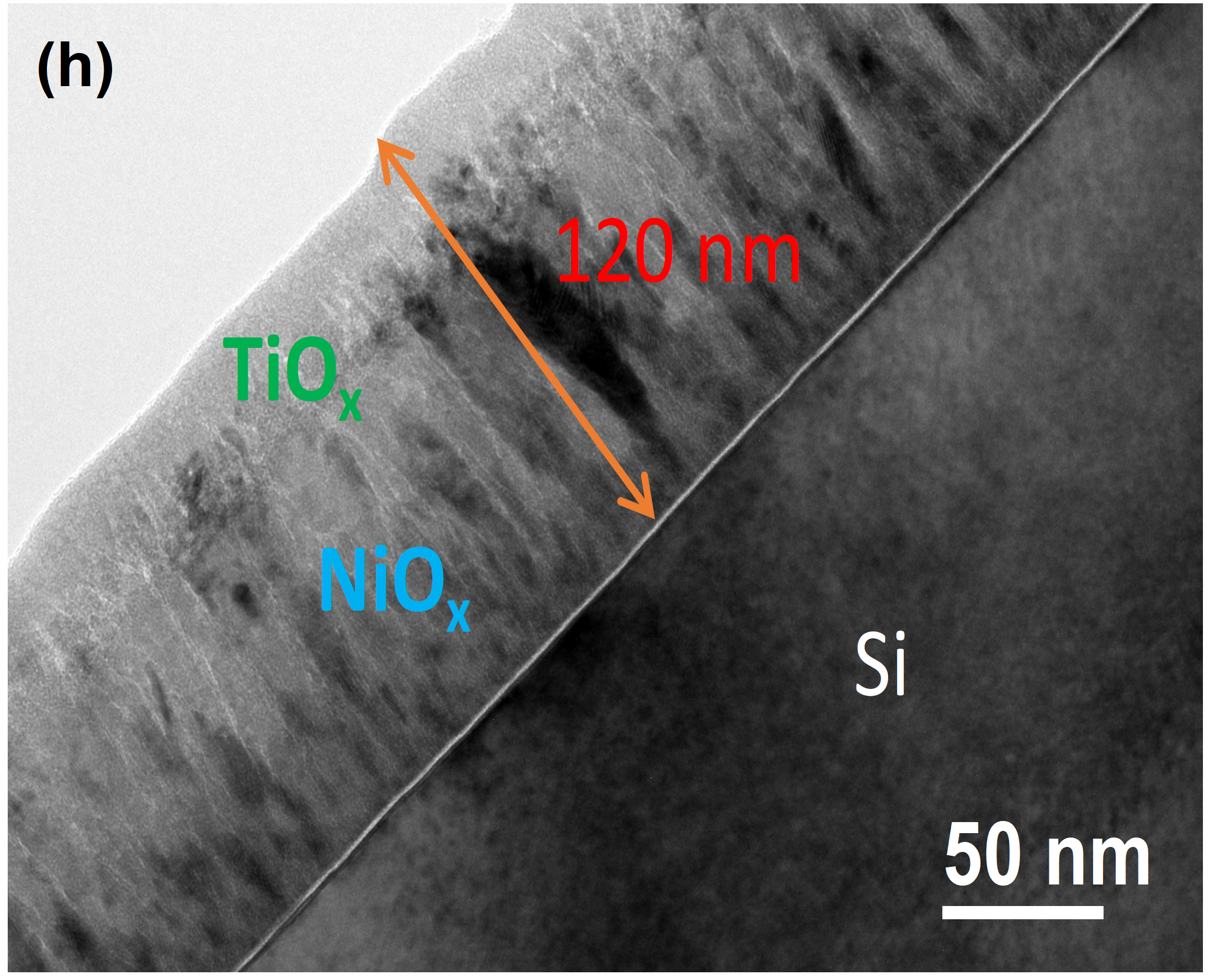}\hfill
\includegraphics[width=0.35\textwidth]{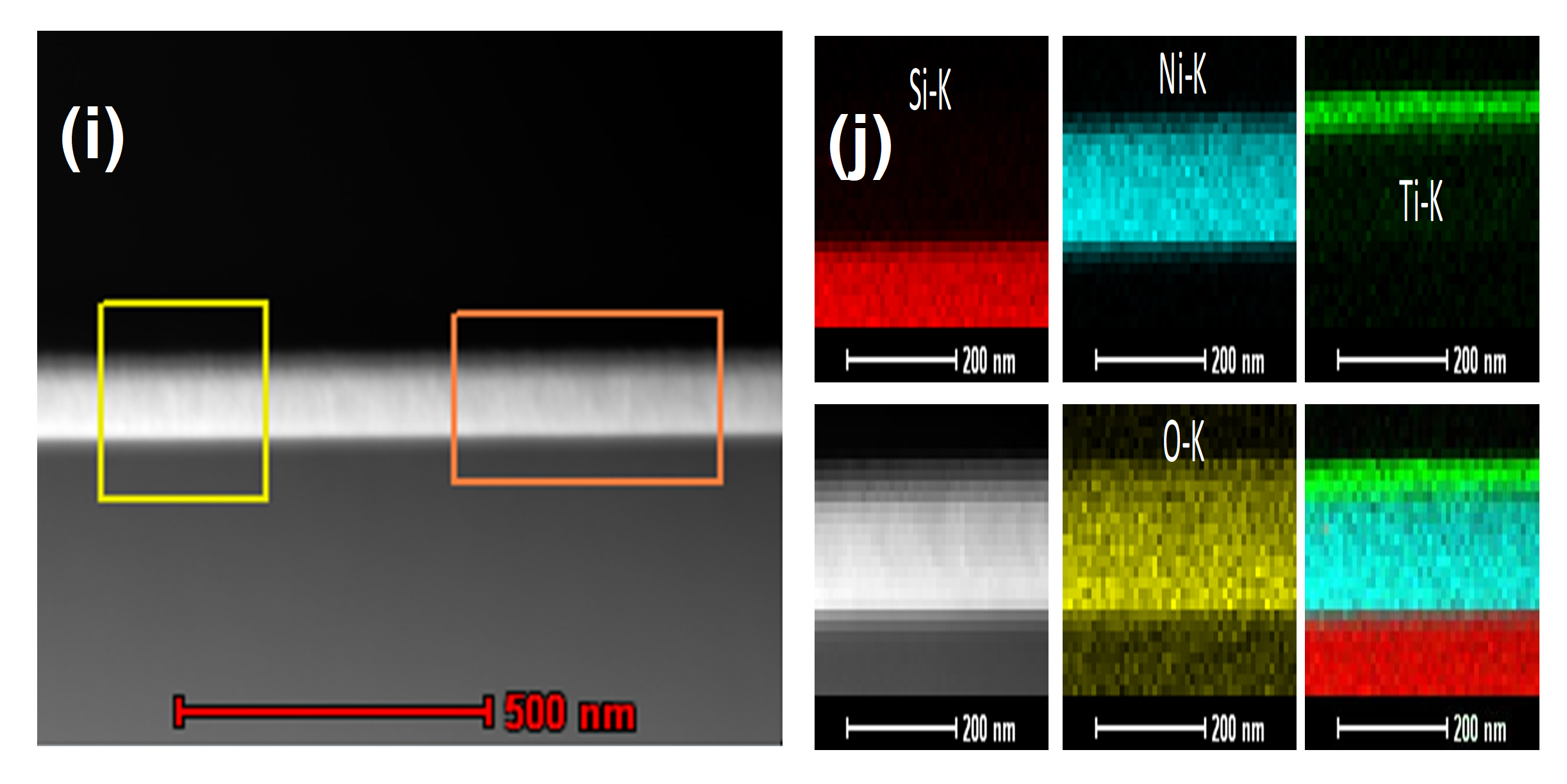}

\caption{Structural, morphological, and chemical characterization of the fabricated TiO$_x$/NiO$_x$ heterostructure: (a) XRD patterns of the NiO$_x$ and TiO$_x$/NiO$_x$ thin films deposited on Si substrates; (b)  SEM image top surface of the TiO$_x$/NiO$_x$ heterostructure; (c) Deconvoluted Ti 2p XPS spectrum and (d) Deconvoluted O 1s XPS spectrum corresponding to the TiO$_x$ surface; (e) Deconvoluted Ni 2p XPS spectrum and (f) Deconvoluted O 1s XPS spectrum corresponding to the NiO$_x$ surface; (g) Low magnification Cross sectional Bright field TEM image of the TiO$_x$/NiO$_x$ heterostructure; (h) HR-TEM image of the heterostructure; (i \& j) STEM-HAADF image and corresponding elemental mapping from orange box region, respectively.}
\label{characterization}
\end{figure*}

Figure \ref{characterization}(a) shows the x-ray diffraction (XRD) patterns of the pristine NiO$_x$ film and the TiO$_x$/NiO$_x$ heterostructure. For the pristine NiO$_x$ film, two distinct diffraction peaks observed at $\sim$ $44^{\circ}$ and $61^{\circ}$ are indexed to (200) and (220) crystallographic planes of NiO, respectively [JCPDS file No:65–2901] confirming the crystalline nature of the NiO film \cite{Do}.  In comparison, the XRD pattern of the TiO$_x$/NiO$_x$ heterostructure retains the characteristic diffraction peaks of NiO indicating that the deposition of the TiO$_x$ layer does not significantly alter the crystalline structure of the underlying NiO layer.  Two new diffraction peaks appeared at $\sim$ $28^{\circ}$ and $43^{\circ}$  are assigned to (110) plane of the anatase phase and (111) plane of the rutile phase, respectively of TiO$_x$ film [JCPDS file No. 89-4920] \cite{Sugapriya}. The coexistence of anatase and rutile related phases suggests the formation of a mixed phase of TiO$_x$ on the NiO$_x$ surface. Thus, the XRD results confirm the successful formation of the TiO$_x$/NiO$_x$ heterostructure while preserving the characteristic crystalline features of both constituent oxide layers.

To investigate the surface morphology of p$^{++}$-Si/NiO$_x$/TiO$_x$ heterostructure, scanning electron microscopy (SEM) was employed. Figure \ref{characterization}(b) shows the top view SEM image of the heterostructure where the inset presents the corresponding particle size distribution obtained using ImageJ software. Statistical analysis reveals an average particle size of about 21 nm. Interestingly, TiO$_x$ thin film deposited directly on the p$^{++}$-Si substrate does not exhibit any distinct particle like morphology  as shown in Figure S1(a). On the contrary, NiO$_x$ thin film deposited on p$^{++}$-Si displays well defined grains with an average size of $\sim$ 7 nm as shown in Figure S1(b). This observation indicates that the surface morphology of the TiO$_x$ layer is strongly influenced by the underlying NiO$_x$ layer. The larger grain size of TiO$_x$ can be attributed to the heterogeneous nucleation and growth of TiO$_x$ on the granular NiO$_x$ surface. Owing to its grainy morphology, the NiO$_x$ layer provides numerous energetically favorable nucleation sites facilitating the initial nucleation of TiO$_x$ and during the film growth, surface diffusion enables adjacent TiO$_x$ nuclei to coalesce resulting in the formation of larger grains \cite{Petrov}. As a result, the average grain size  of the top TiO$_x$ layer increases to about 21 nm. These observations demonstrate that the introduction of the NiO$_x$ interlayer significantly modifies the growth kinetics of the TiO$_x$ film resulting in an enlarged grain  and altered surface morphology. Figure S2(a) and S2(b) present two dimensional (2D) and three dimensional (3D) AFM topographical images of the heterostructure, respectively. The AFM analysis reveals a smooth and uniform surface morphology with a root mean square (RMS) surface roughness of approximately 1.25 nm. The low RMS roughness indicates the formation of a compact and homogeneous TiO$_x$ top layer with minimal surface irregularities.

To investigate the chemical composition and oxidation states of the constituent layers, x-ray photoelectron spectroscopy (XPS) measurements were carried out. Initially, XPS analysis was performed on the p$^{++}$-Si/NiO$_x$/TiO$_x$ sample with the same layer thicknesses also maintained for the devices. The survey spectrum and high-resolution spectra clearly reveal the presence of Ti, O and adventitious C. However, no detectable Ni signal was observed here. The absence of the Ni signal is attributed to the surface sensitive nature of XPS technique. The measurements were performed using a monochromatic Al K$_\alpha$ x-ray source ($h\nu = 1486.6$~eV) for which the photoelectron escape depth is typically limited to  a few nanometers only. Since the TiO$_x$ layer is the topmost layer having a thickness of 30 nm, the underlying NiO$_x$ layer could not be probed. To determine the chemical states of the NiO$_x$ layer, XPS measurement was carried out on the NiO$_x$ thin film of identical thickness as in heterostructure, deposited on a p$^{++}$-Si substrate under identical deposition conditions and elemental composition and oxidation states were identified. Figure \ref{characterization}(c) presents the deconvoluted high-resolution Ti 2$p$ XPS spectrum of the TiO$_x$ layer. The spectral deconvolution was performed using CasaXPS software. Prior to fitting, the binding energy scale was calibrated considering the adventitious C 1$s$ peak to 284.6 eV. A GL(30) line shape combined with a Shirley background was employed for peak fitting. During the fitting procedure, the spin--orbit splitting between the Ti 2$p_{3/2}$ and Ti 2$p_{1/2}$ components was fixed at 5.6 eV while the corresponding peak area ratio was constrained to 2:1. The Ti 2$p$ spectrum can be resolved into two oxidation states. The peaks located at 458.40 and 464.0 eV are assigned to the Ti$^{3+}$ 2$p_{3/2}$ and Ti$^{3+}$ 2$p_{1/2}$ components, respectively whereas the peaks centered at 459.4 and 465.0 eV correspond to the Ti$^{4+}$ 2$p_{3/2}$ and Ti$^{4+}$ 2$p_{1/2}$ states \cite{Dip}. The coexistence of Ti$^{4+}$ and Ti$^{3+}$ oxidation states indicates partial reduction of Ti which is widely recognized as a signature of oxygen vacancy formation in TiO$_x$ film. Figure \ref{characterization}(d)  shows the high resolution O 1$s$ XPS spectrum of the TiO$_x$ film. The deconvoluted spectrum consists of two distinct components. The dominant peak centered at 530.1 eV is attributed to lattice oxygen (O$^{2-}$) bonded to Ti atoms within the TiO$_x$ crystal framework. In contrast, the higher binding energy component located at 531.3 eV is assigned to non-lattice oxygen species which are commonly associated with oxygen deficient regions and oxygen vacancies in TiO$_x$ \cite{Ham, Ju}. The presence of this higher binding energy peak along with Ti$^{3+}$ component observed in the Ti 2$p$ spectrum is a strong evidence of the existence of oxygen vacancies in the deposited TiO$_x$ thin film.
Figure \ref{characterization}(e) displays the deconvoluted high resolution Ni 2$p$ XPS spectrum of NiO$_x$ thin film. The spectrum was analyzed using CasaXPS software in a similar fashion. The spin orbit splitting between the Ni 2$p_{3/2}$ and Ni 2$p_{1/2}$ components was constrained to 17.3 eV whereas the area ratio of the doublet was fixed at 2:1 throughout the fitting process. The fitted spectrum consists of contributions from both Ni$^{2+}$ and Ni$^{3+}$ oxidation states. The peaks located at 855.2 and 872.5 eV are assigned to the Ni$^{2+}$ 2$p_{3/2}$ and Ni$^{2+}$ 2$p_{1/2}$ components, respectively whereas the peaks centered at 856.9 and 874.2 eV correspond to the Ni$^{3+}$ 2$p_{3/2}$ and Ni$^{3+}$ 2$p_{1/2}$ states \cite{Cheng}. Two additional broad features observed at higher binding energies are attributed to the characteristic shake up satellite peaks of nickel oxide. The simultaneous presence of Ni$^{2+}$ and Ni$^{3+}$ confirms that the deposited NiO$_x$ film is non-stoichiometric in nature. Figure \ref{characterization}(f) shows the deconvoluted high resolution O 1$s$ XPS spectrum of the NiO$_x$ thin film. The spectrum is composed of three distinct components. The peak located at 529.5 eV is assigned to lattice oxygen (O$^{2-}$) wheras another peak centered at 531.4 eV originates from non-lattice oxygen species associated with oxygen vacancy induced defects and oxygen deficient regions in the film. The third component, observed at 533.2 eV is attributed to weakly bound oxygen species in the form of interstitial oxygen and surface hydroxyl groups \cite{Chen, Ran}. The XPS analysis confirms the presence of oxygen vacancy related defect states in both TiO$_x$ and NiO$_x$ layers of the heterostructure.

Figures \ref{characterization}(g) and (h) show the cross-sectional TEM image of the fabricated p$^{++}$-Si/NiO$_x$/TiO$_x$ heterostructure. The HR-TEM images confirm the formation of a well defined bilayer structure comprising NiO$_x$ and TiO$_x$ layers of $\sim$ 90 nm and $\sim$ 30 nm thickness, respectively with a total film thickness of about 120 nm . The high angle annualar dark-field scanning TEM (HAADF-STEM) image of the heterostructure and the corresponding Energy dispersive X-ray Spectroscopy (EDX) elemental mapping presented in Figure \ref{characterization}(i) and (j) also confirm the presence of Si, Ni, Ti and O.
 
\subsection{Resistive Switching Characteristics of the Ag/TiO$_x$/NiO$_x$/p$^{++}$-Si Device}
\begin{figure*}[t]
\centering
\includegraphics[width=0.32\textwidth]{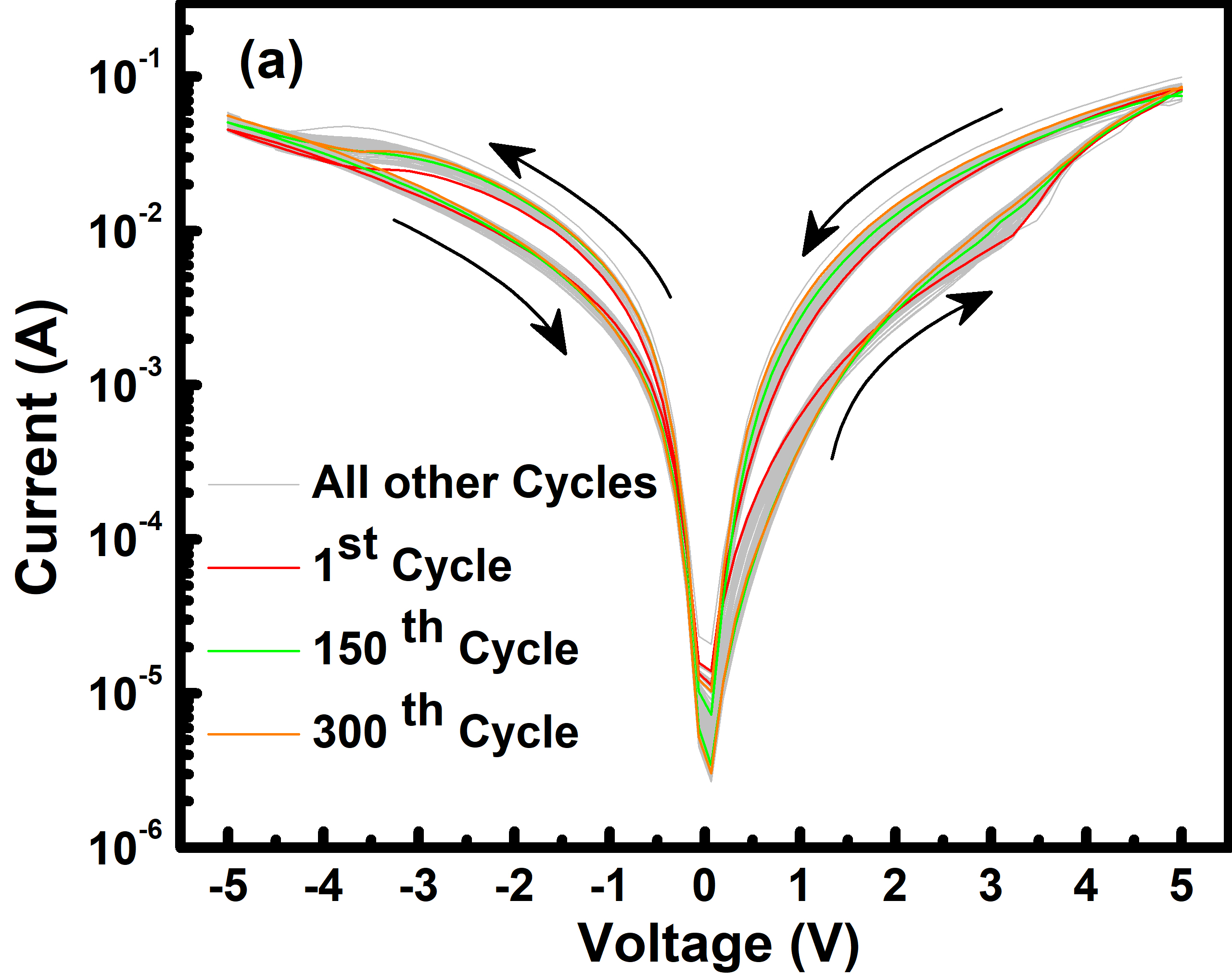}\hfill
\includegraphics[width=0.34\textwidth]{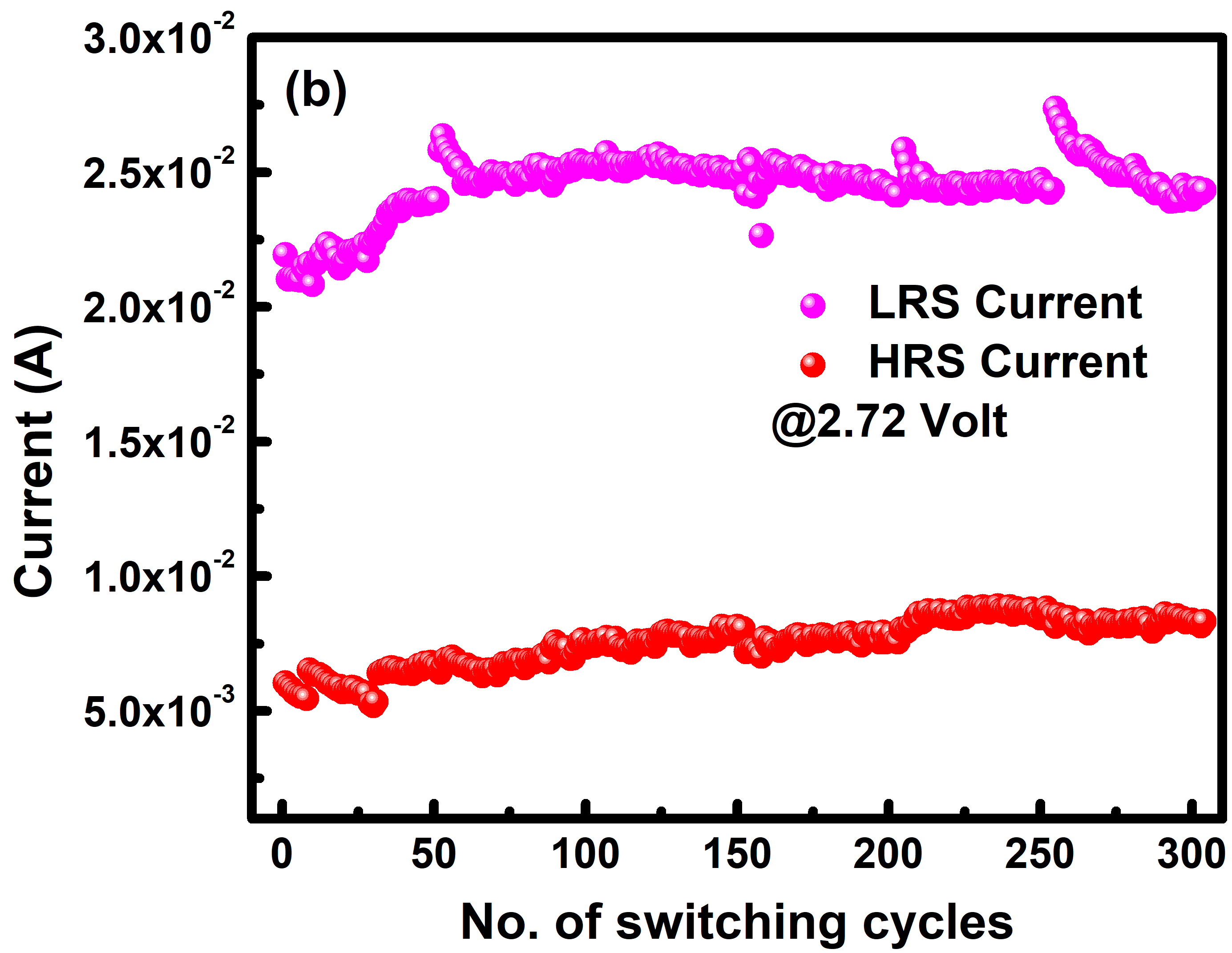}\hfill
\includegraphics[width=0.34\textwidth]{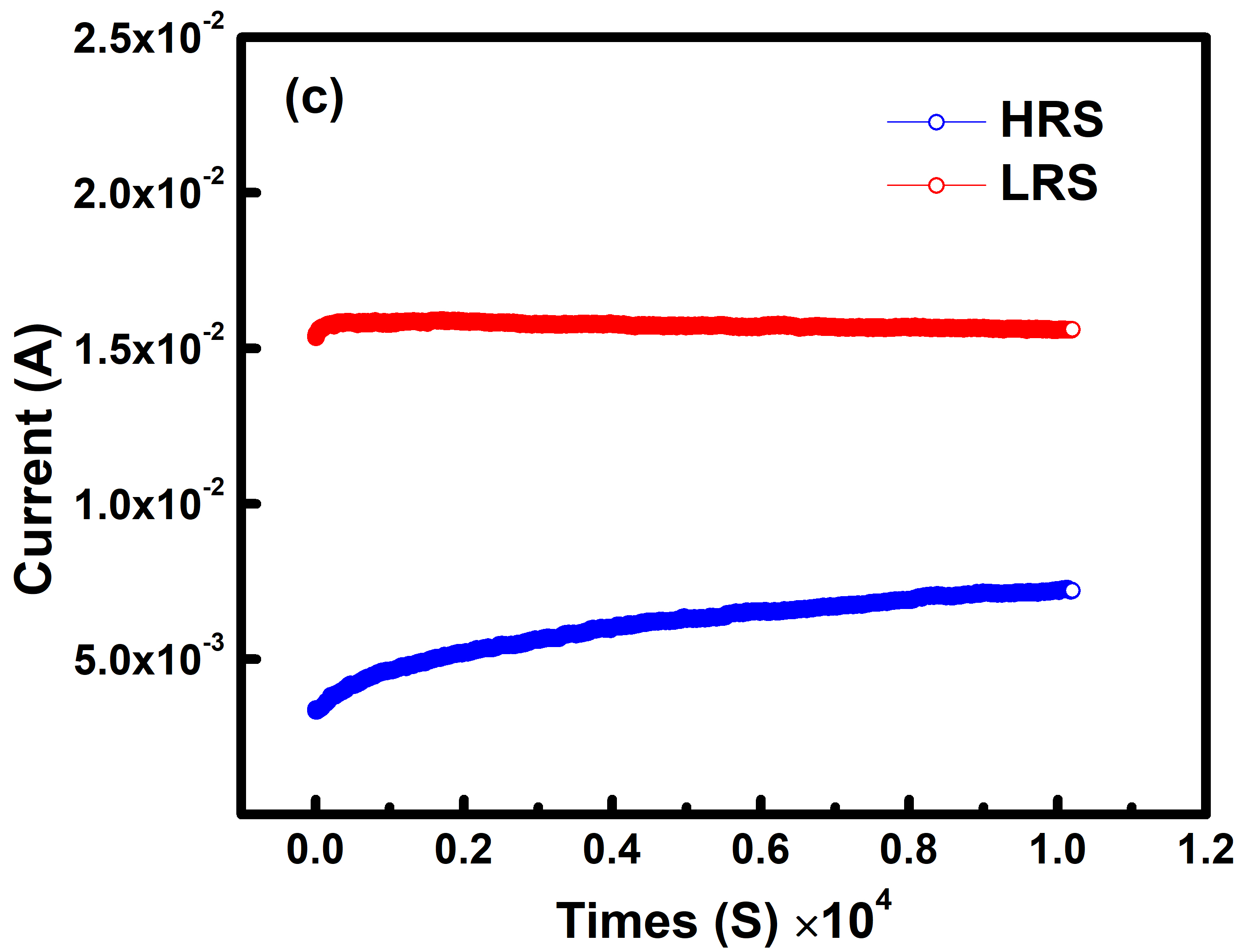}

\vspace{5mm}

\includegraphics[width=0.34\textwidth]{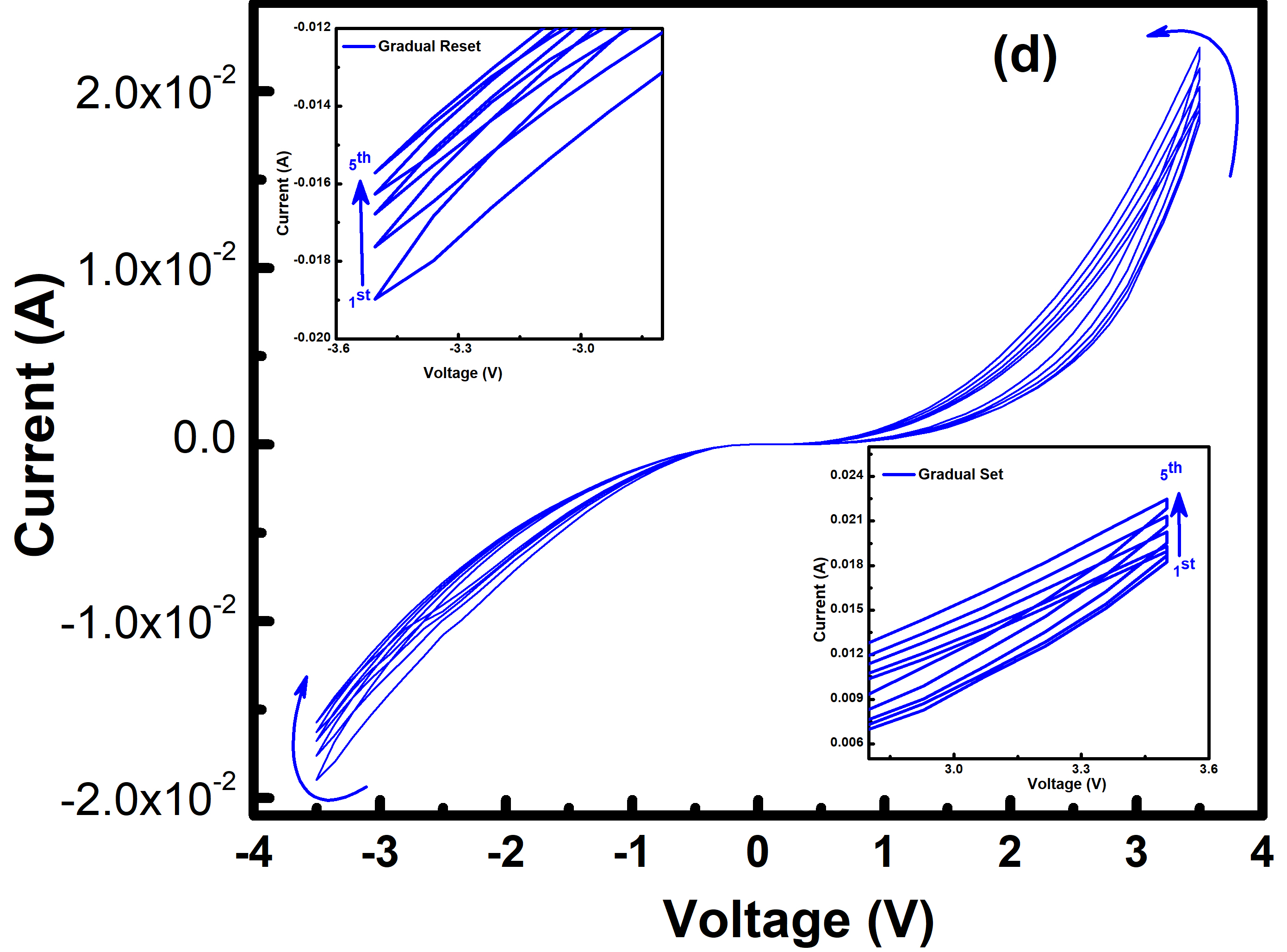}\hfill
\includegraphics[width=0.32\textwidth]{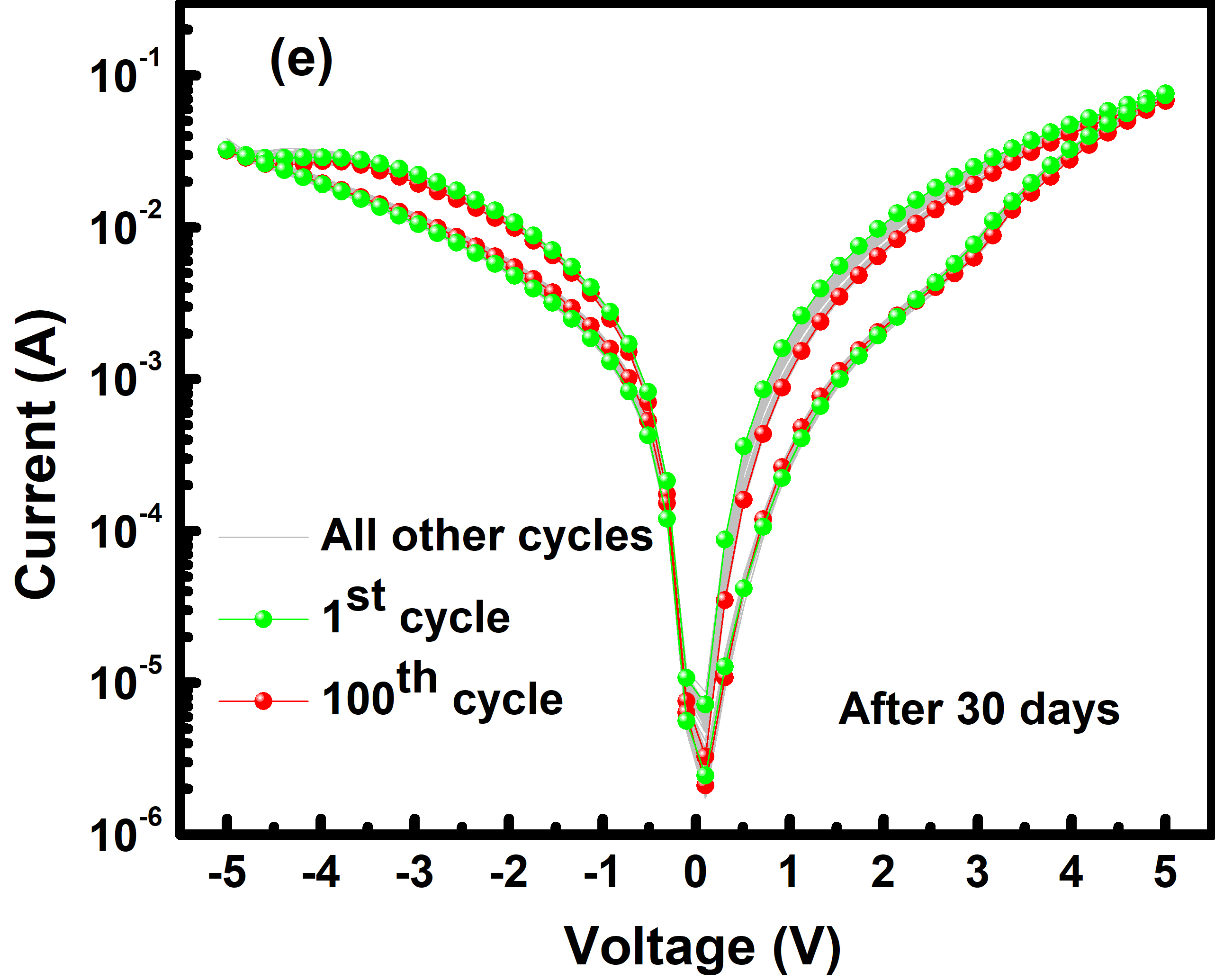}\hfill
\includegraphics[width=0.32\textwidth]{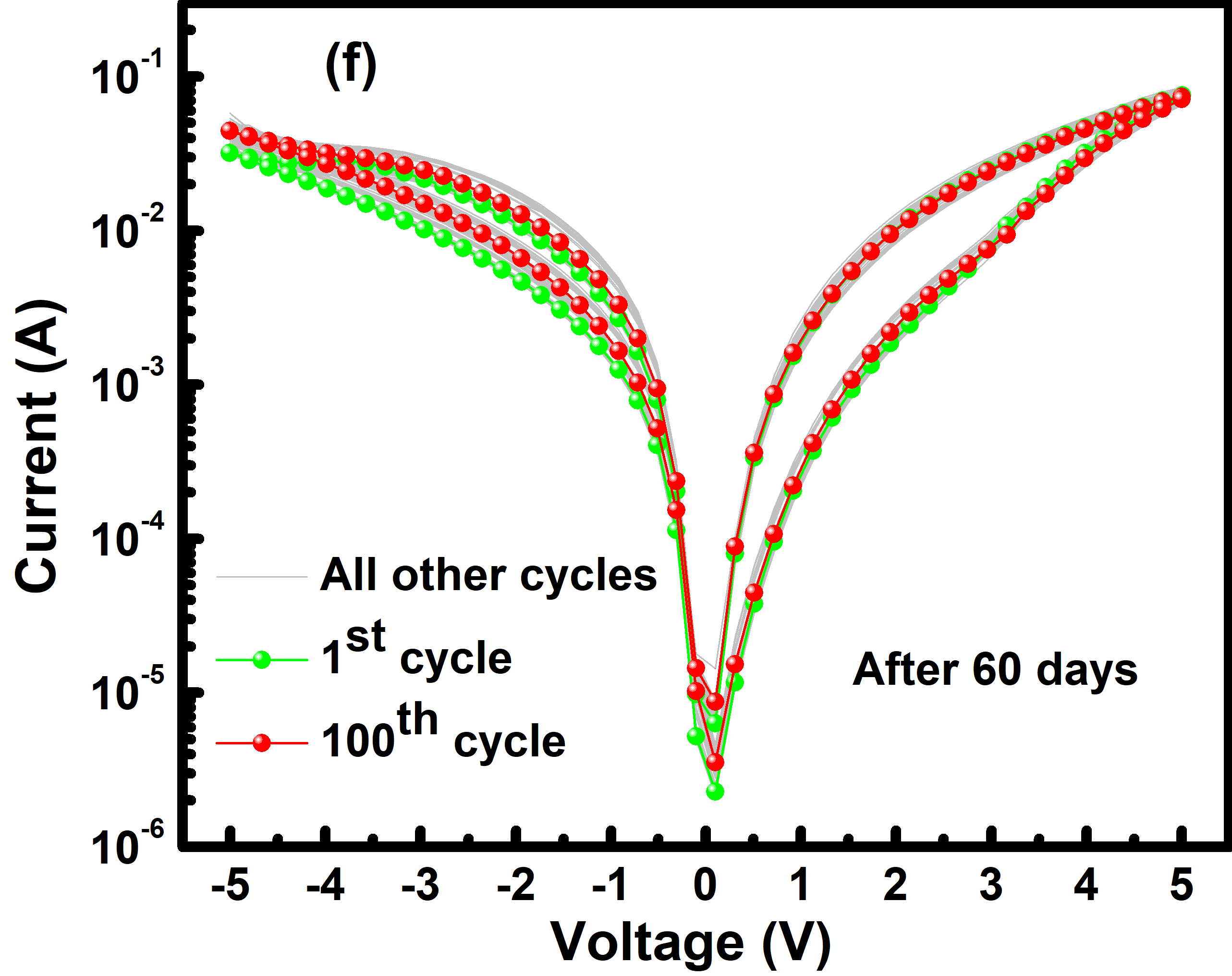}

\vspace{5mm}
\includegraphics[width=0.34\textwidth]{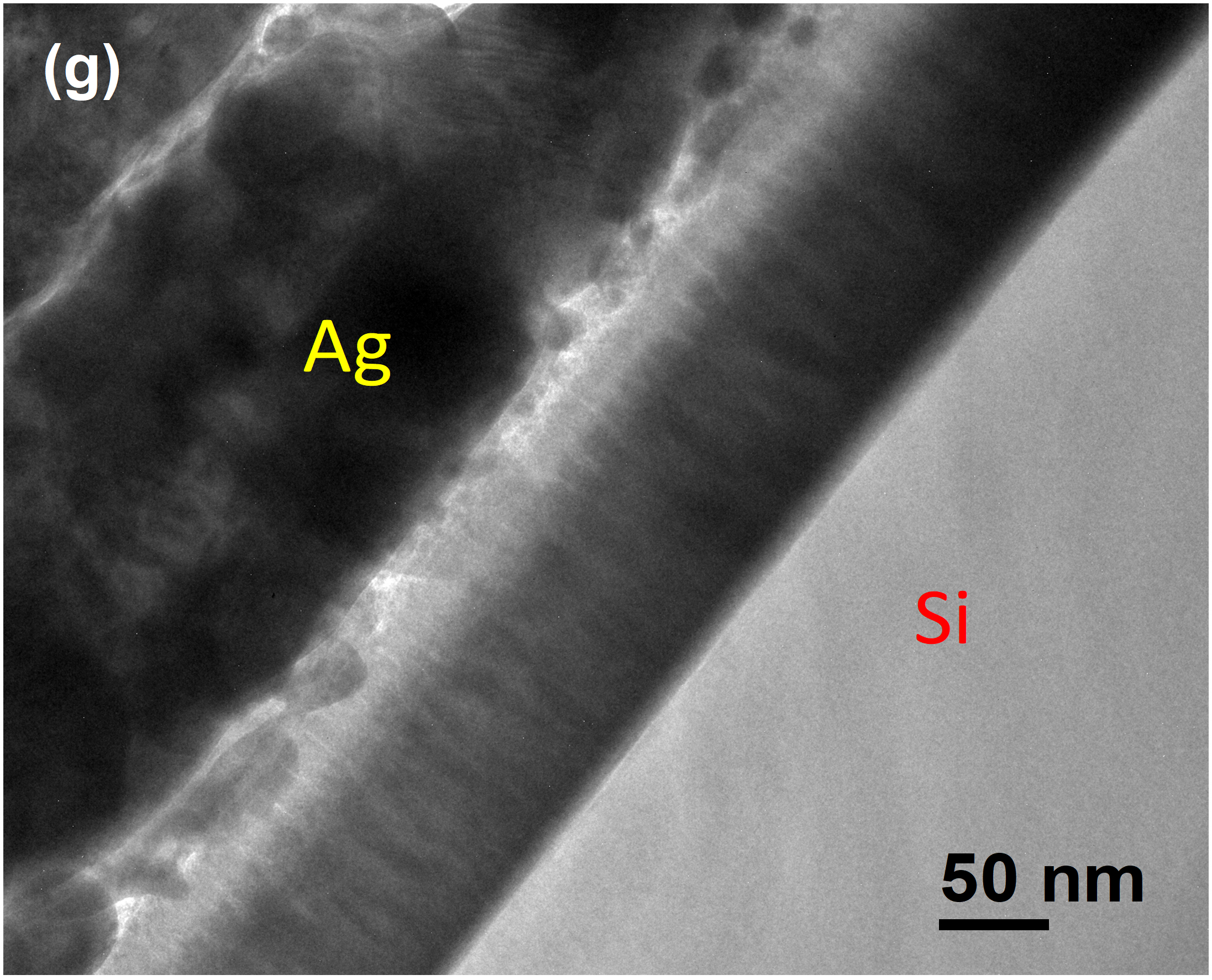}
\hspace{5mm}
\includegraphics[width=0.34\textwidth]{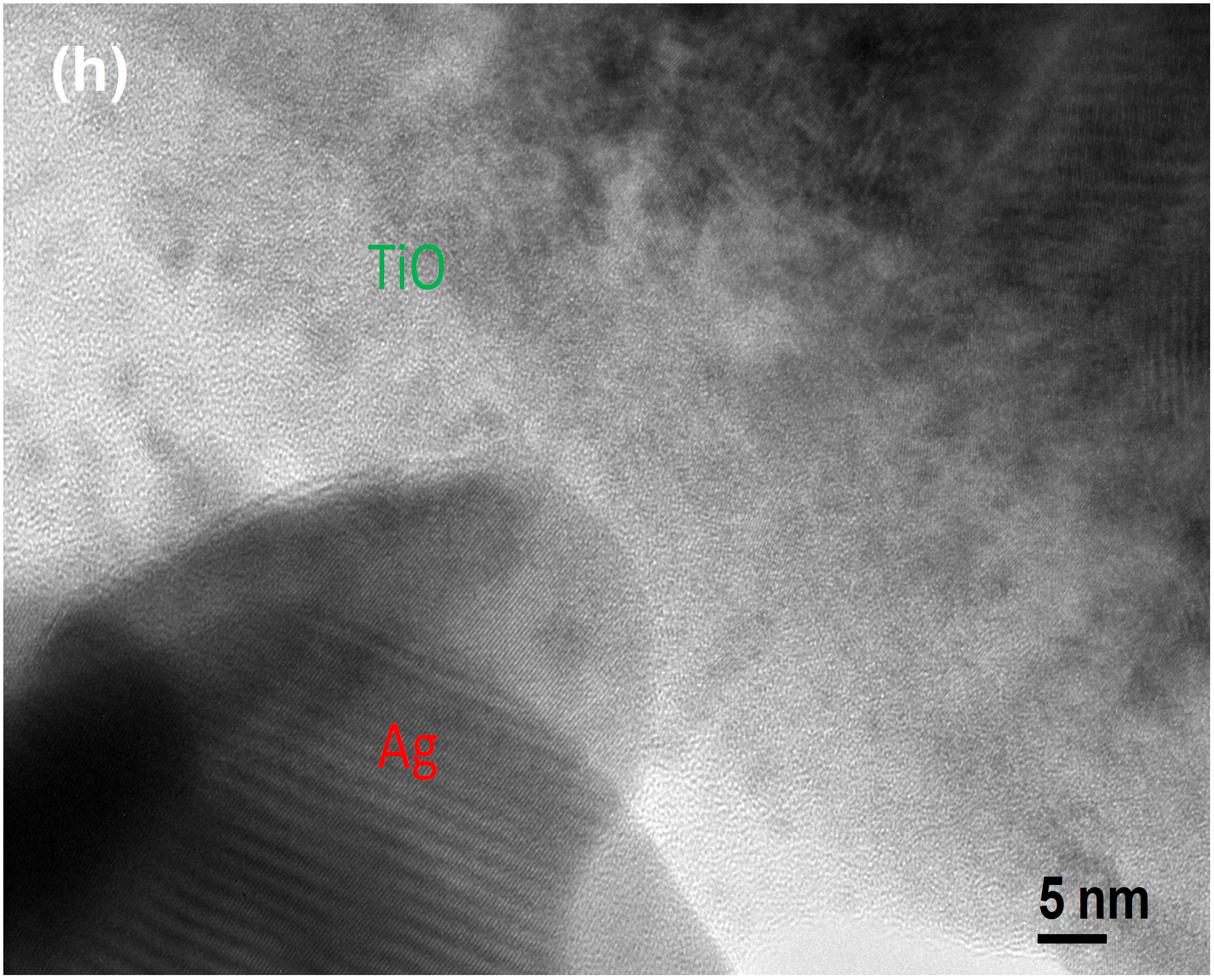}

\caption{Electrical characteristics and long term stability of the TiO$_x$/NiO$_x$ heterostructure memristor: (a) I--V characteristics of the fabricated device; (b) Endurance characteristics over repeated switching cycles; (c) Retention characteristics of the resistance states; (d) Gradual SET and RESET characteristics demonstrating analog resistance modulation; (e \& f) I--V characteristics measured after 30 and 60 days, respectively, demonstrating the long-term stability of the device; (g) TEM image of the memristor after electrical measurements; and (h) HR-TEM image showing Ag migration after electrical measurements.}
\label{I-V}
\end{figure*}
The electrical performance of the Ag/TiO$_x$/NiO$_x$/p$^{++}$-Si device was evaluated upon application of a continuous DC voltage sweep to the Ag top electrode connecting the p$^{++}$-Si bottom to ground terminal. The voltage was varied following the sequence $0 \rightarrow V_{\mathrm{max}} \rightarrow 0 \rightarrow -V_{\mathrm{max}} \rightarrow 0$ to investigate the resistive switching behaviour. The  current--voltage (I--V) characteristics of the memristor are presented in Figure S3(a). Figure \ref{I-V}(a) presents the current--voltage ($I$--$V$) characteristics for 300 consecutive switching cycles with a current compliance limit of 100 mA. The highly reproducible hysteretic response confirms distinct bipolar resistive switching without any preliminary electroforming process.

During the positive voltage sweep, the current initially changes at a slower rate and afterwards starts varying rapidly at about +2.32 V which is identified as the SET voltage ($V_{\mathrm{SET}}$).  The slight variation in $V_{\mathrm{SET}}$ throughout repeated DC sweeps indicates uniform switching behaviour. Under negative voltage bias, the device progressively returns from the LRS to the HRS. The corresponding RESET transition occurs at approximately -2.50 V. The opposite polarities required for the SET and RESET processes confirm the bipolar nature of the switching behaviour. The endurance performance of the device over 300 consecutive switching cycles is presented in Figure \ref{I-V}(b). The current values corresponding to the LRS and HRS were extracted at a read voltage of 2.72 V. Both the states remain distinguishable throughout the repeated switching operation with limited cycle to cycle fluctuations demonstrating good endurance and operational stability. Figure \ref{I-V}(c) presents the retention characteristics recorded for $1.1\times 10^{4}$ s at a constant read voltage of 2.0 V. The LRS and HRS currents remain sufficiently separated over the entire measurement duration preserving an $I_{\mathrm{ON}}/I_{\mathrm{OFF}}$ ratio of $10^{1}$. The stable retention of the programmed resistance states together with the relatively low operating voltages demonstrates the reliability of the device for nonvolatile memory applications. Figure \ref{I-V}(d) shows five consecutive $I$--$V$ sweeps measured within the voltage range of $+3.5$ to $-3.5$ V. During the positive sweep, the current progressively increases from the first cycle to the fifth cycle revealing a gradual SET process. In contrast, the current continuously decreases during the negative bias sweep indicating a gradual RESET transition. The progressive evolution of the device current demonstrates that its resistance can be modulated continuously rather than through an abrupt transition between two discrete resistance states. Such analog switching behaviour is advantageous for implementing tunable synaptic weights in neuromorphic computing systems. The long term reproducibility of the resistive switching response was studied by repeating the electrical measurements on the same device after storage periods of 30 and 60 days. Figures \ref{I-V}(e) and (f) show 100 consecutive $I$--$V$ switching cycles obtained after 30 and 60 days, respectively. The characteristic bipolar switching behaviour remains well preserved even after 60 days with a minor current variation. These findings demonstrate that the device maintains a stable electrical performance over an extended period. Device to device reproducibility was further assessed using four independently fabricated devices from the same batch. The inset of Figure S3(a) shows current--voltage (I--V) characteristics of the fabricated memristor. The device exhibits asymmetric current conduction with a maximum current of approximately 10 mA during the positive voltage sweep and 20 mA during the negative voltage sweep indicating a rectifying behaviour. The observed rectification originates primarily from the $p-n$ heterojunction formed between the $p$-type NiO$_x$ and $n$-type TiO$_x$ layers. During the positive voltage sweep, the heterojunction is reverse biased resulting in a suppressed current. In contrast, the negative voltage sweep forward biases the junction leading to a relatively higher current. The representative $I$--$V$ characteristics of the other three devices are presented in Figures S3(b)--(d) of the Supporting Information. Comparable bipolar resistive switching characteristics were observed for all investigated devices indicating good fabrication repeatability and consistent electrical performance across the device batch.  Furthermore, as the highly doped p$^{++}$-Si substrate possesses a very low resistivity, the p$^{++}$-Si/NiO$_x$ contact is considered to be nearly ohmic.

We also studied the $I-V$ characteristics of the Ag/NiO$_x$/p$^{++}$-Si device as shown in Figure S4. The device exhibits clear resistive switching behaviour but the memory window becomes gradually narrower with increasing switching cycles indicating poor endurance and instability of the switching characteristics. In contrast, the Ag/TiO$_x$/NiO$_x$/p$^{++}$-Si device exhibits significantly improved switching stability. This enhanced stability can be attributed to the formation of the TiO$_x$/NiO$_x$ heterojunction which provides an additional interface for regulating the transport and redistribution of oxygen vacancies during the switching process. Thus, the heterojunction plays a crucial role in improving the endurance and stability of the Ag/TiO$_x$/NiO$_x$/p$^{++}$-Si device.

After completing all the electrical measurements, cross-sectional TEM analysis was again performed on the same device. Figure \ref{I-V}(g) shows the HR-TEM image of the electrically tested p$^{++}$-Si/NiO$_x$/TiO$_x$/Ag device. As highlighted in the magnified image, several Ag atoms are observed to have diffused into the TiO$_x$ layer and is presented in Figure \ref{I-V}(h). Further, the elemental mapping of the cross section is depicted in Figure S5. The diffusion of Ag into the TiO$_x$ layer is attributed to the electric-field-driven migration of Ag ions during conductive filament formation. Under an applied bias, Ag atoms of the top electrode are oxidized and migrate through the TiO$_x$ layer where they are reduced to metal Ag atom. These Ag metal atoms effectively enhancing the built up electric field across the junction contribute to the filament formation within the depletion layer.

\subsection{Artificial Synaptic behaviour of the Ag/TiO$_x$/NiO$_x$/p$^{++}$-Si Memristor}
\begin{figure*}[!t]
\centering

\subfloat[]{
\includegraphics[width=0.92\textwidth]{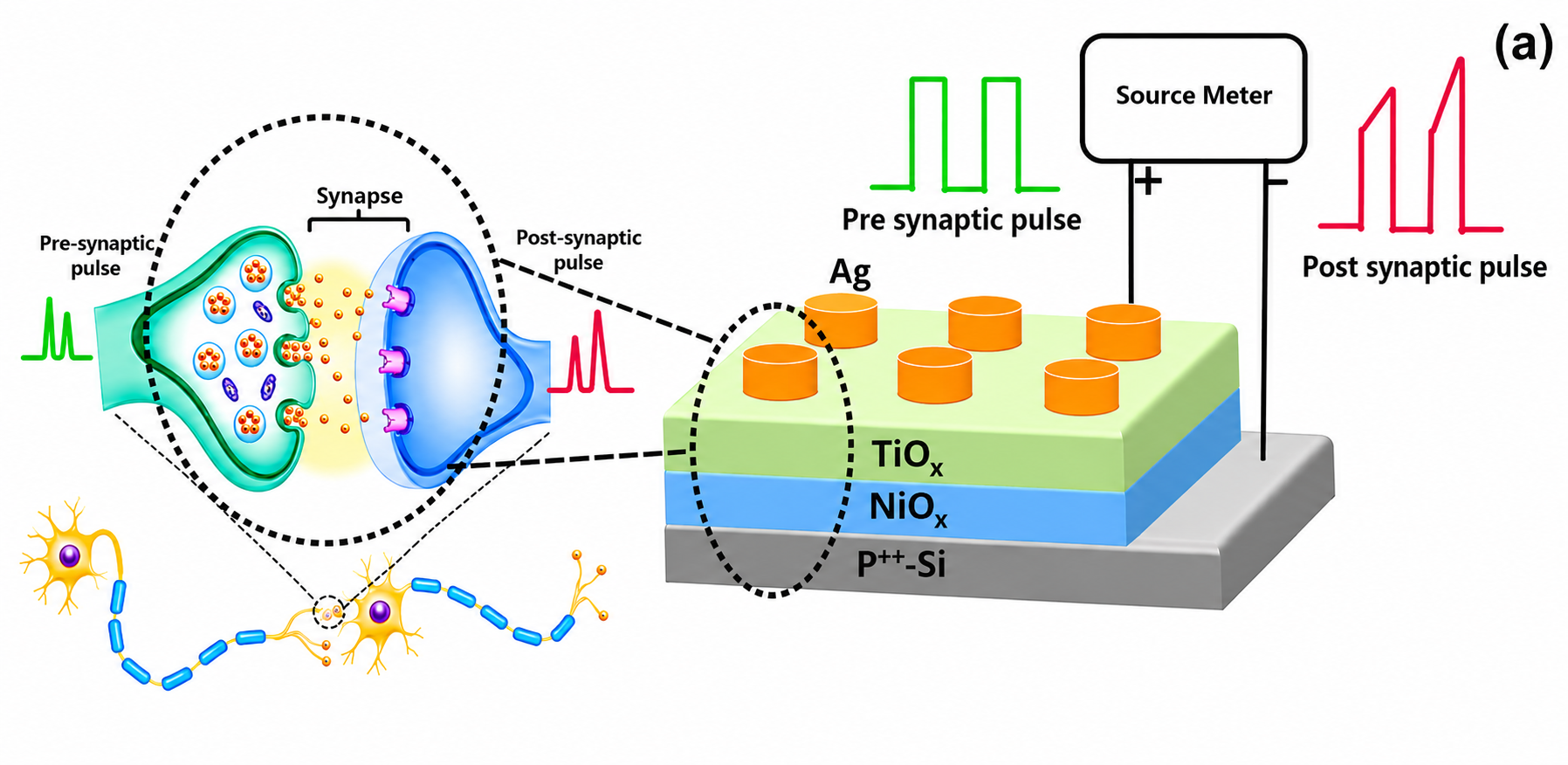}
}

\vspace{0.3cm}

\subfloat[]{
\includegraphics[width=0.31\textwidth]{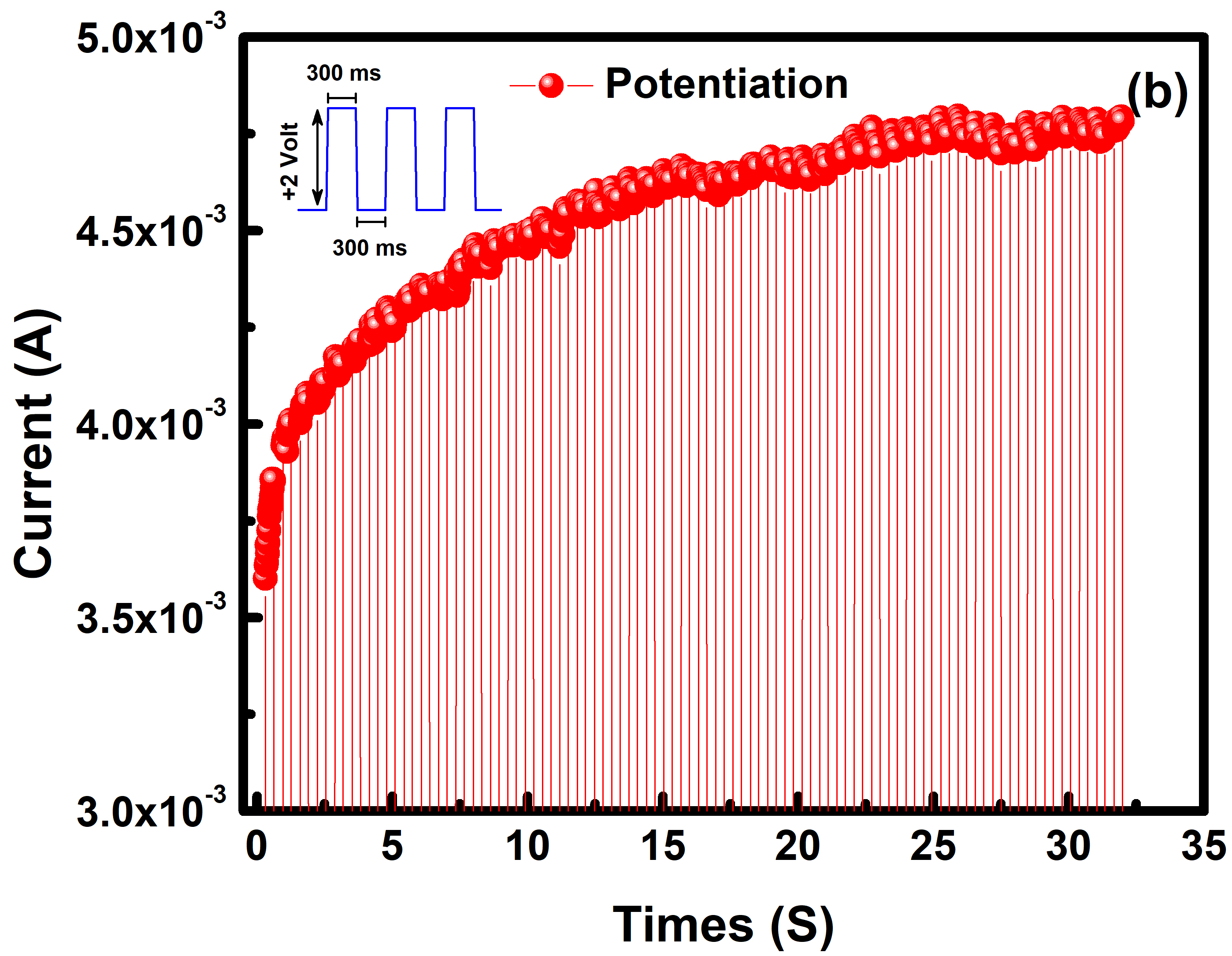}
}
\hfill
\subfloat[]{
\includegraphics[width=0.31\textwidth]{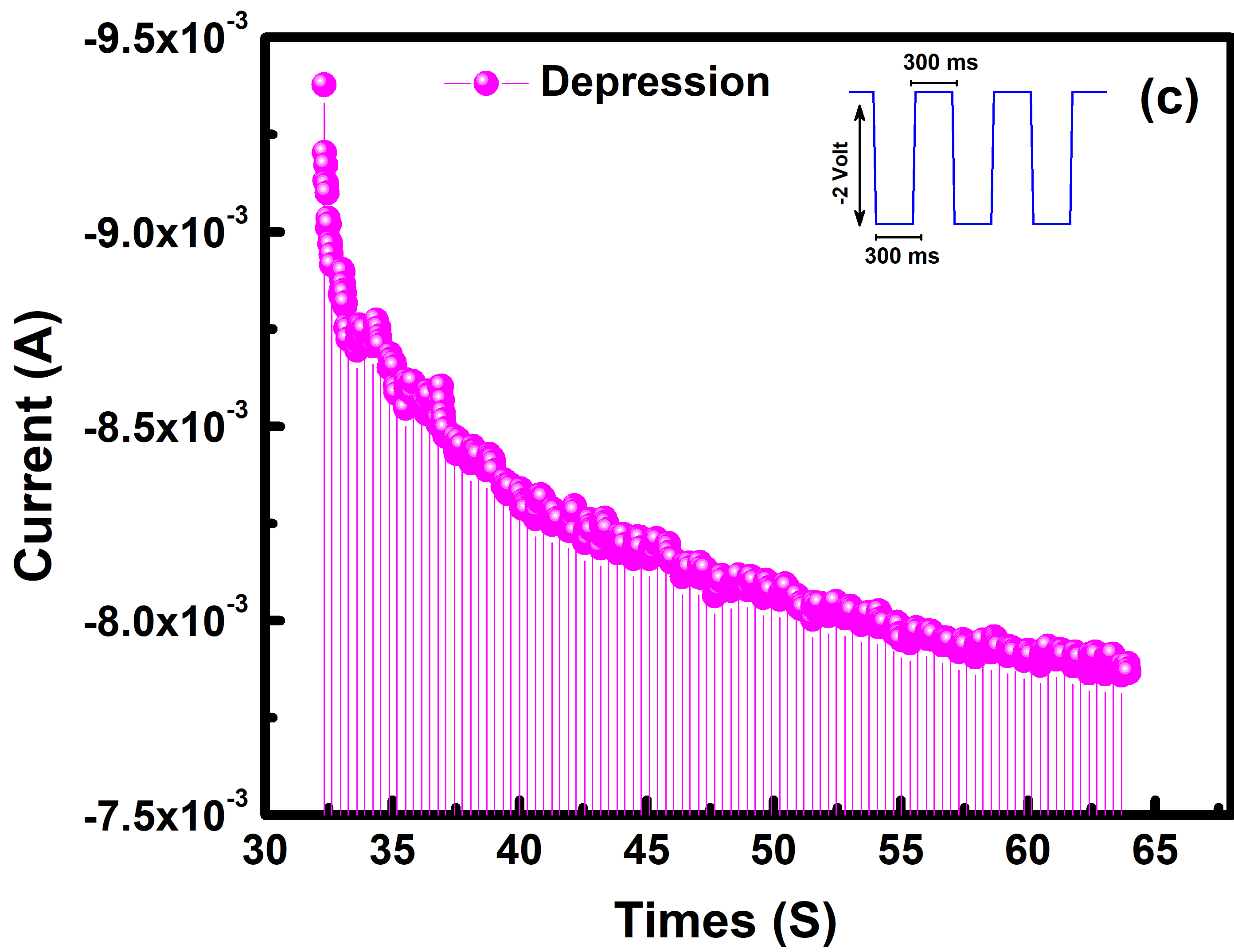}
}
\hfill
\subfloat[]{
\includegraphics[width=0.31\textwidth]{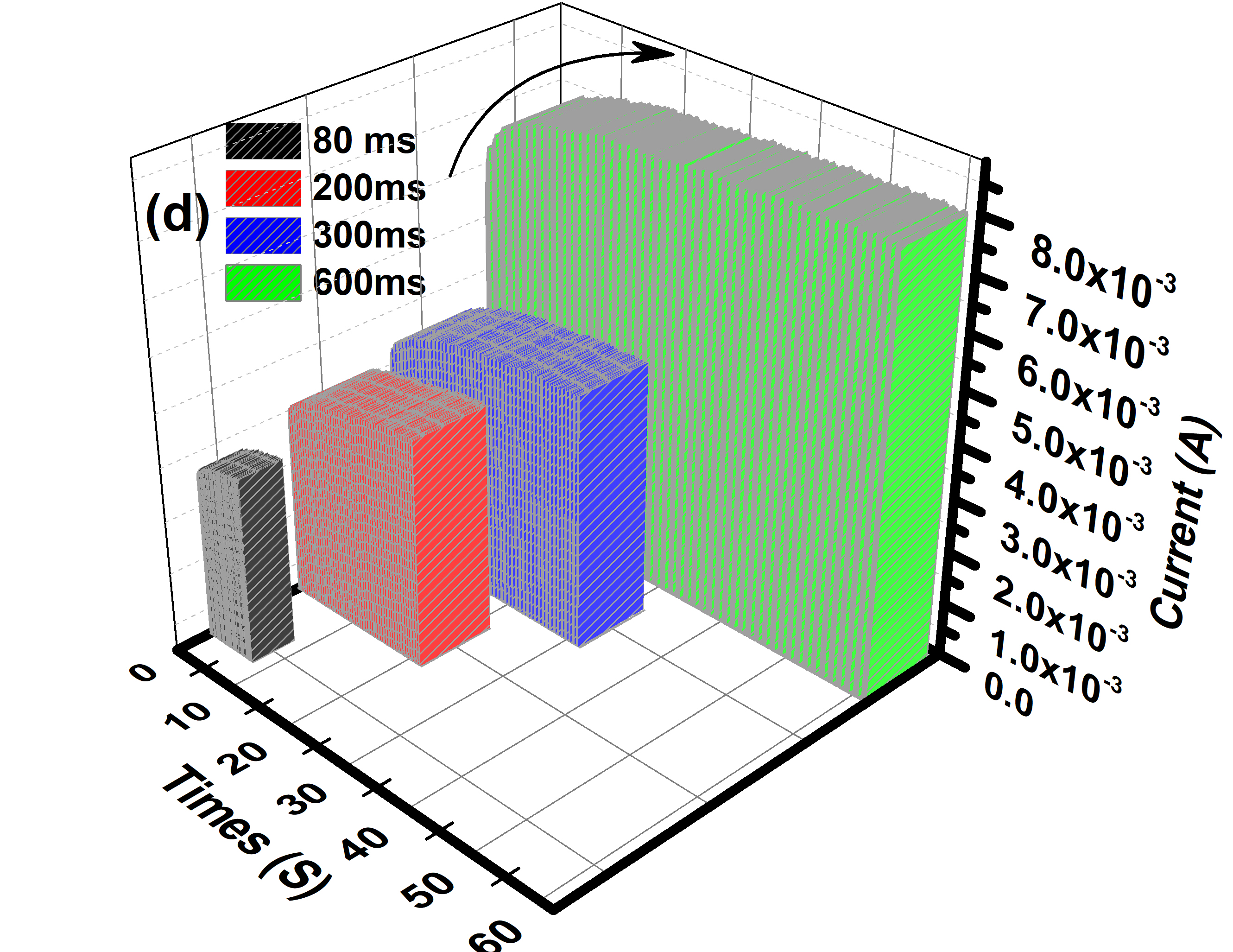}
}

\vspace{0.3cm}

\subfloat[]{
\includegraphics[width=0.31\textwidth]{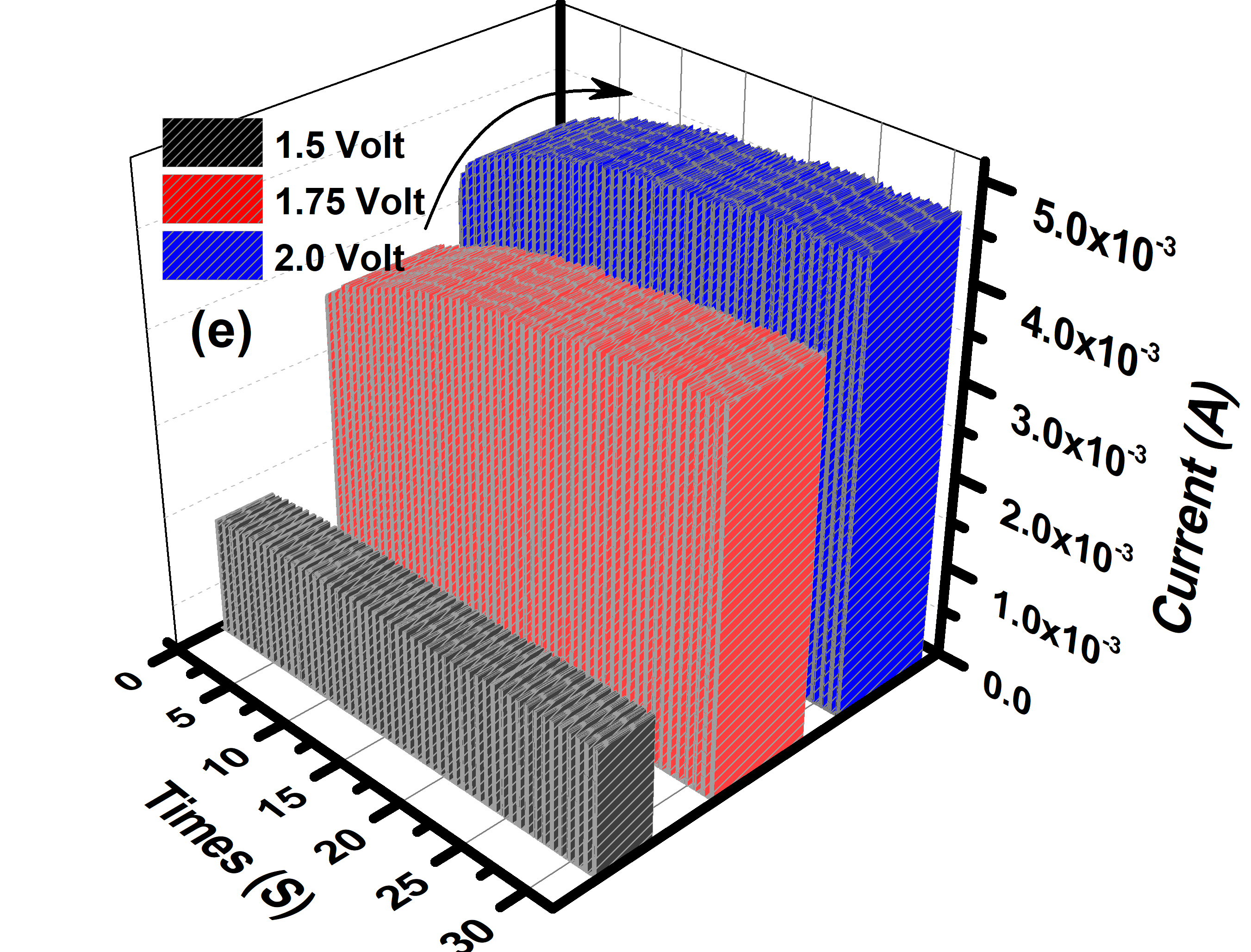}
}
\hfill
\subfloat[]{
\includegraphics[width=0.35\textwidth]{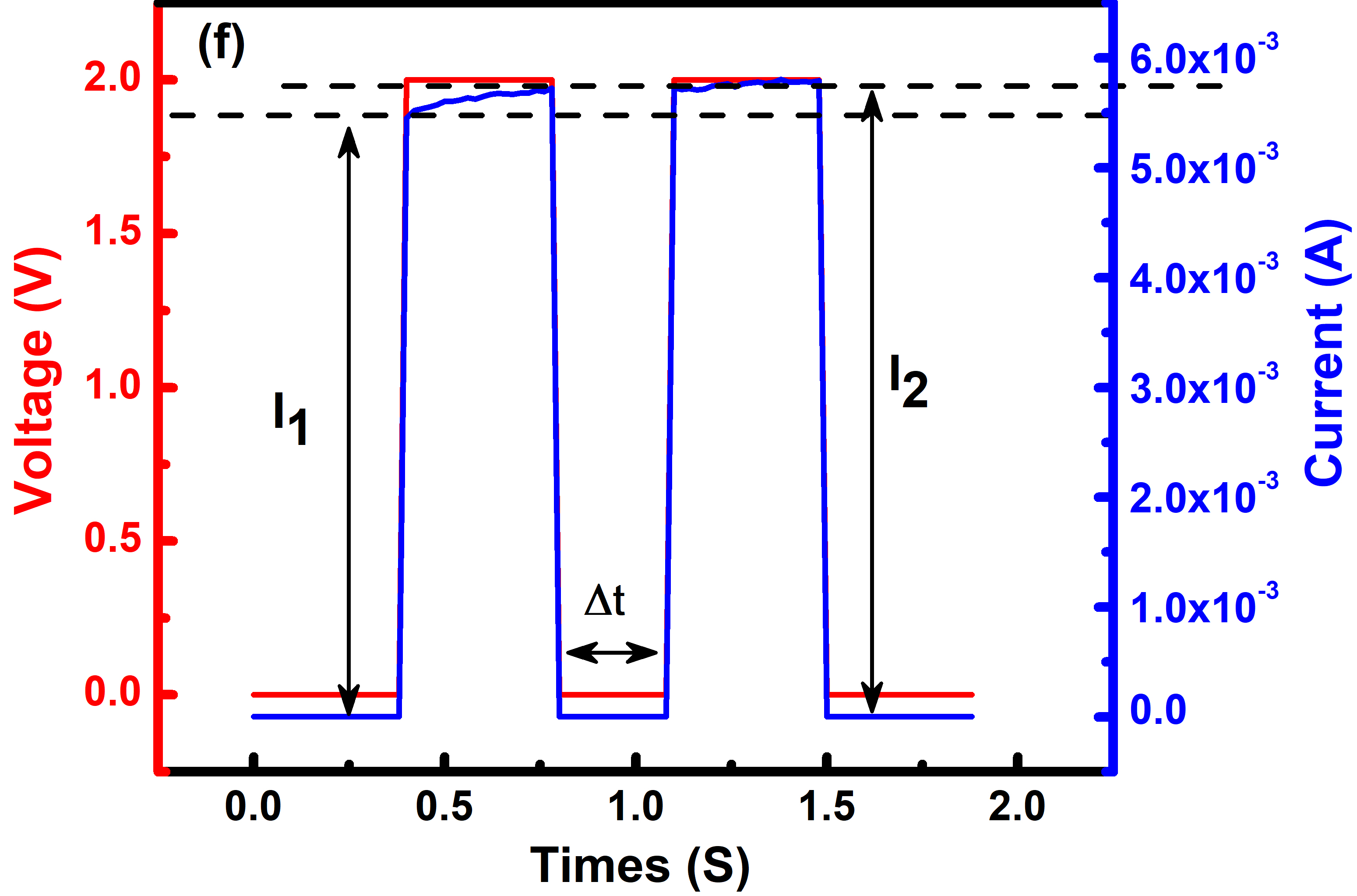}
}
\hfill
\subfloat[]{
\includegraphics[width=0.31\textwidth]{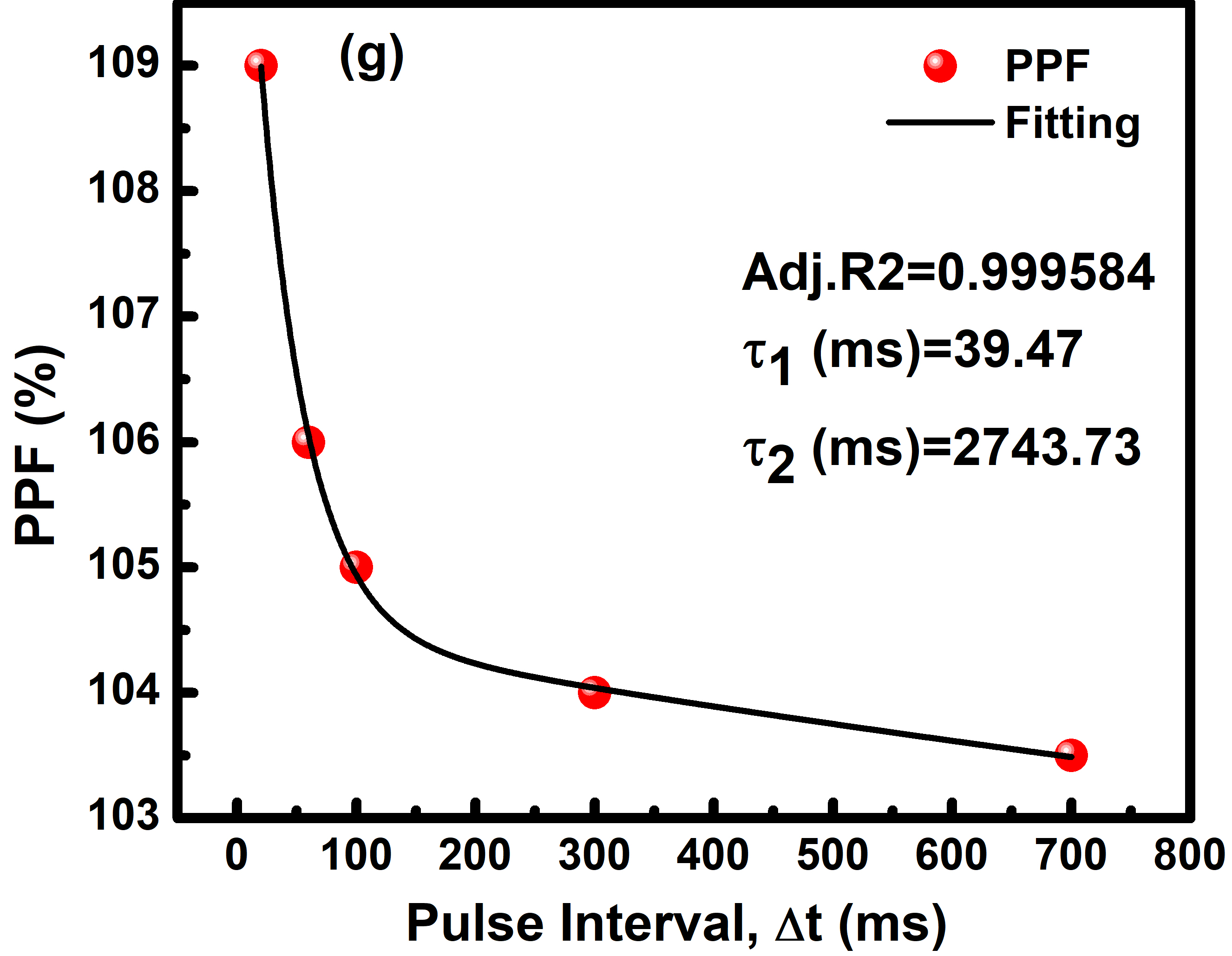}
}

\caption{Synaptic response of  Ag/TiO$_x$/NiO$_x$/p$^{++}$-Si Memristive Device: (a) A schematic illustration comparing the biological synapse with the Ag/TiO$_x$/NiO$_x$/p$^{++}$-Si memristor emphasizing its bio-inspired synaptic architecture and operation; (b) Analog potentiation and (c) Analog depression characteristics of the memristor achieved after 60 consecutive programming pulses where the inset shows the corresponding pulse waveform parameters; (d) The current response measured by varying the pulse width from 80 to 600 ms maintaining a constant pulse amplitude of 2 V; (e) The current response recorded under voltage pulses with amplitudes varying from 1.5 to 2.0 V maintaining a pulse width of 300 ms and an interpulse interval of 300 ms; (f) Response current and applied voltage pulses for paired pulse facilitation (PPF); (g) Relationship between the PPF index and the interpulse interval.}
\label{synaptic}
\end{figure*}

Figure \ref{I-V}(d) shows SET and RESET switching characteristics of the memristor where a continuous change in conductance is observed. Such a gradual change enables the device to emulate the behaviour of biological synapses for neuromorphic computing applications. In biological neural systems, synaptic plasticity is governed by the strengthening (potentiation) and weakening (depression) the synaptic connections in response to external stimuli. These adaptive changes play an important role in learning and memory and provide the basic foundation for the development of artificial neural networks \cite{Zh, Wang}. In biological neural systems, neurotransmitters released from the presynaptic neuron trigger excitatory or inhibitory responses in the postsynaptic neuron giving rise to excitatory postsynaptic current (EPSC) or inhibitory postsynaptic current (IPSC), respectively. Repeated excitatory or inhibitory stimulation gradually modifies the synaptic efficacy forming the basis of learning and memory through synaptic plasticity. To investigate the artificial synaptic behaviour of the memristor, a sequence of electrical voltage pulses was applied to the device. Figure \ref{synaptic}(a) illustrates a schematic of the biological synaptic process underlying signal transmission between neurons, which is emulated by the Ag/TiO$_x$/NiO$_x$/P$^{++}$-Si memristor as an artificial synaptic device. Figures \ref{synaptic}(b) and (c) present the long term potentiation (LTP) and long term depression (LTD) characteristics at 2.0 V while maintaining a pulse width and interpulse interval of 300 ms. During the potentiation process, the device conductance increases progressively under consecutive positive voltage pulses whereas an opposite trend is observed under negative voltage pulses resulting in gradual conductance suppression during depression. The analog modulation of the conductance demonstrates stable and repeatable synaptic weight updates closely resembling the learning and forgetting processes observed in biological synapses. The dependence of the synaptic response on pulse width is presented in Figure \ref{synaptic}(d). The potentiation characteristics were investigated by applying 60 consecutive positive voltage pulses with a fixed amplitude of 2.0 V while varying the pulse width from 80 to 600 ms. The device exhibites a progressively incremental current response with increasing duty cycle indicating enhanced synaptic weight modulation under prolonged electrical stimulation. Furthermore, the influence of pulse amplitude on synaptic potentiation is also illustrated in Figure \ref{synaptic}(e) where 60 identical voltage pulses with amplitudes of 1.5, 1.75 and 2.0 V were applied while maintaining a constant pulse width of 300 ms. A gradual increase in the response current is observed with increasing pulse amplitude demonstrating that stronger electrical stimuli produce more pronounced potentiation in the Ag/TiO$_x$/NiO$_x$/p$^{++}$-Si artificial synapse.

In the biological system, paired pulse facilitation (PPF), is usually defined as a momentary increase in the possibility of vesicular release due to already accumulated residual Ca$^{2+}$ at the synaptic terminal and a lasting effect of Ca$^{2+}$ on the exocytotic Ca$^{2+}$  sensor at the time of application of the second stimulus \cite{Alex}. The fast slope occurs in the PPF vs time plot at very brief inter pulse intervals typically raging between 50--100 ms evidenced from a steep slope as the residual free calcium ions (Ca$^{2+}$) from the first pulse accumulate rapidly right next to the vesicular release sensors \cite{Chessa}. The slow slope taken place at longer time intervals of hundreds of ms to sec is characterized by a gradual, shallow slope as the rapid calcium component decays away. These Ca$^{2+}$ ions act as the primary trigger for neurotransmitter release by binding to specialized vesicular release sensors in the presynaptic nerve terminal. As the specialized vesicular release sensors contain aspartate, strongly electronegative oxygen atoms in -COO$^-$ groups, the Ca$^{2+}$ ions form an ionic or coordinate covalent interaction with the carboxylate site. 

Similar trend is observed in case of an artificial synapse. Here, PPF is manifested by a larger excitatory postsynaptic current (EPSC) generated by the second stimulus compared with the preceding one when the interpulse interval is shorter than the relaxation time of the device. Figure \ref{synaptic}(f) presents the applied paired voltage pulses together with the corresponding current response of the Ag/TiO$_x$/NiO$_x$/p$^{++}$-Si memristor. Two consecutive voltage pulses with an amplitude of 2.0 V with a pulse width of 300 ms and interpulse interval of 300 ms were applied to the device. As evidenced from the figure, the current generated by the second pulse is noticeably higher than that generated by the first pulse demonstrating the occurrence of paired pulse facilitation. The dependence of the PPF index on the interpulse interval is presented in Figure \ref{synaptic} (g). As the interval between the two pulses increases, the PPF index gradually decreases. The experimental data can be well described by a double exponential decay function: \cite{Zucker}
\begin{equation}
\mathrm{PPF}(\%)=\frac{I_{2}}{I_{1}}\times100
=1+C_{1}e^{-\Delta t/\tau_{1}}
+C_{2}e^{-\Delta t/\tau_{2}}
\label{PPF}
\end{equation}
where $I_1$ and $I_2$ represent the peak current responses to the first and second pulses, respectively, $\Delta t$ is the interpulse interval, $C_1$ and $C_2$ are fitting coefficients, and $\tau_1$ and $\tau_2$ denote the characteristic relaxation time constants associated with the fast and slow relaxation processes of the artificial synapse. The fitted relaxation time constants $\tau_{1}$ and $\tau_{2}$ represent the fast and slow decay processes associated with paired pulse facilitation, respectively. Specifically, $\tau_{1}=39.47$ ms characterizes the rapid relaxation component whereas $\tau_{2}=2.74$ s corresponds to the slower relaxation process. The fitting coefficients $C_{1}$ and $C_{2}$ quantify the initial facilitation contributions of the fast and slow components, respectively. The main charged species in oxides responsible for filament formation and lower resistance state are accumulation/alignment of ionized oxygen vacancy defects in presence of a voltage pulse. In absence of the pulse, as a natural consequence, the alignment breaks and the charged species tend to diffuse back over time to their thermal equilibrium sites \cite{Sungjun}. However, the spontaneous polarization of oxygen depleted from NiO$_x$-TiO$_x$ interface/layer hinders this process resulting in a significant vacancy removal responsible for filament formation causing an abrupt reduction in the device conductivity having a decay constant of $\tau_1$. The remaining vacancies present disappear slowly with another time constant $\tau_2$. This transient decay fits with double exponential functions whose time constants depend on the local concentration, migration energy and spatial distribution gradient of the driving vacancies. The above dynamics of the process is responsible for the current decay with two slopes after cessation of the pulse. When a new pulse is applied after the former one, the conduction front moves forward due to persistence of polarization and shorter pulse interval resulting in an increase of the current. 

\begin{figure*}[!t]
\centering

\includegraphics[width=0.55\textwidth]{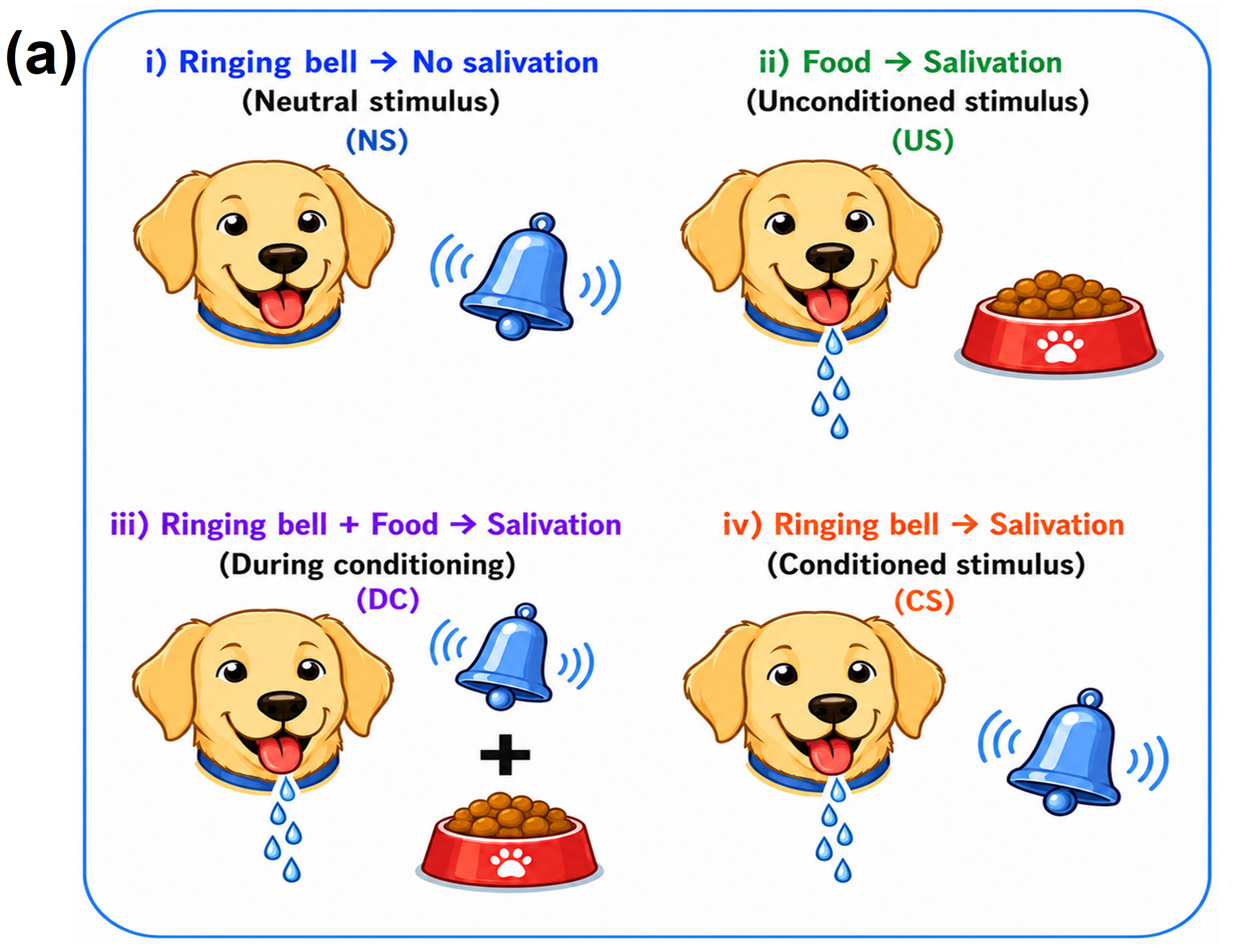}

\vspace{0.5cm}

\includegraphics[width=0.475\textwidth]{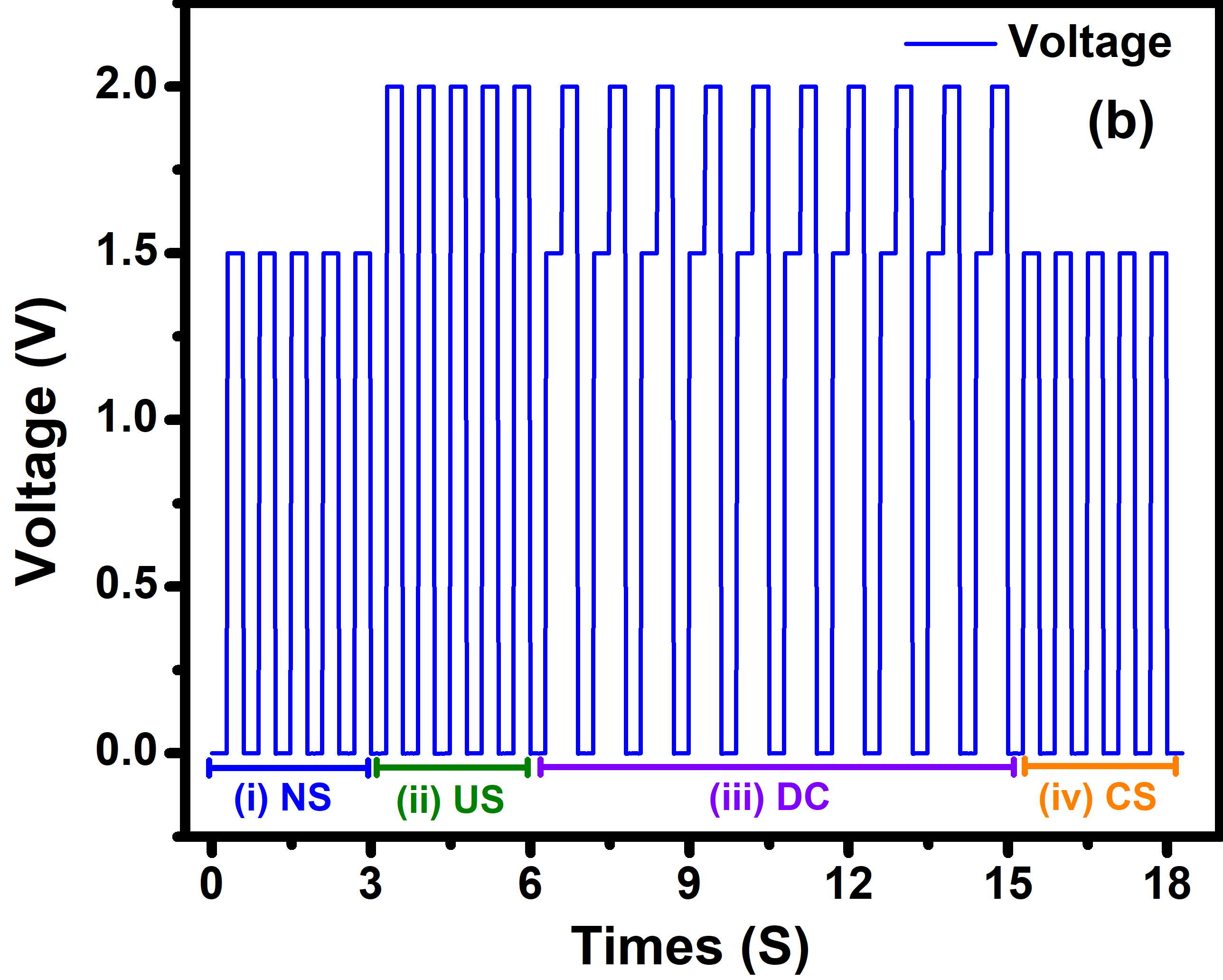}
\hfill
\includegraphics[width=0.51\textwidth]{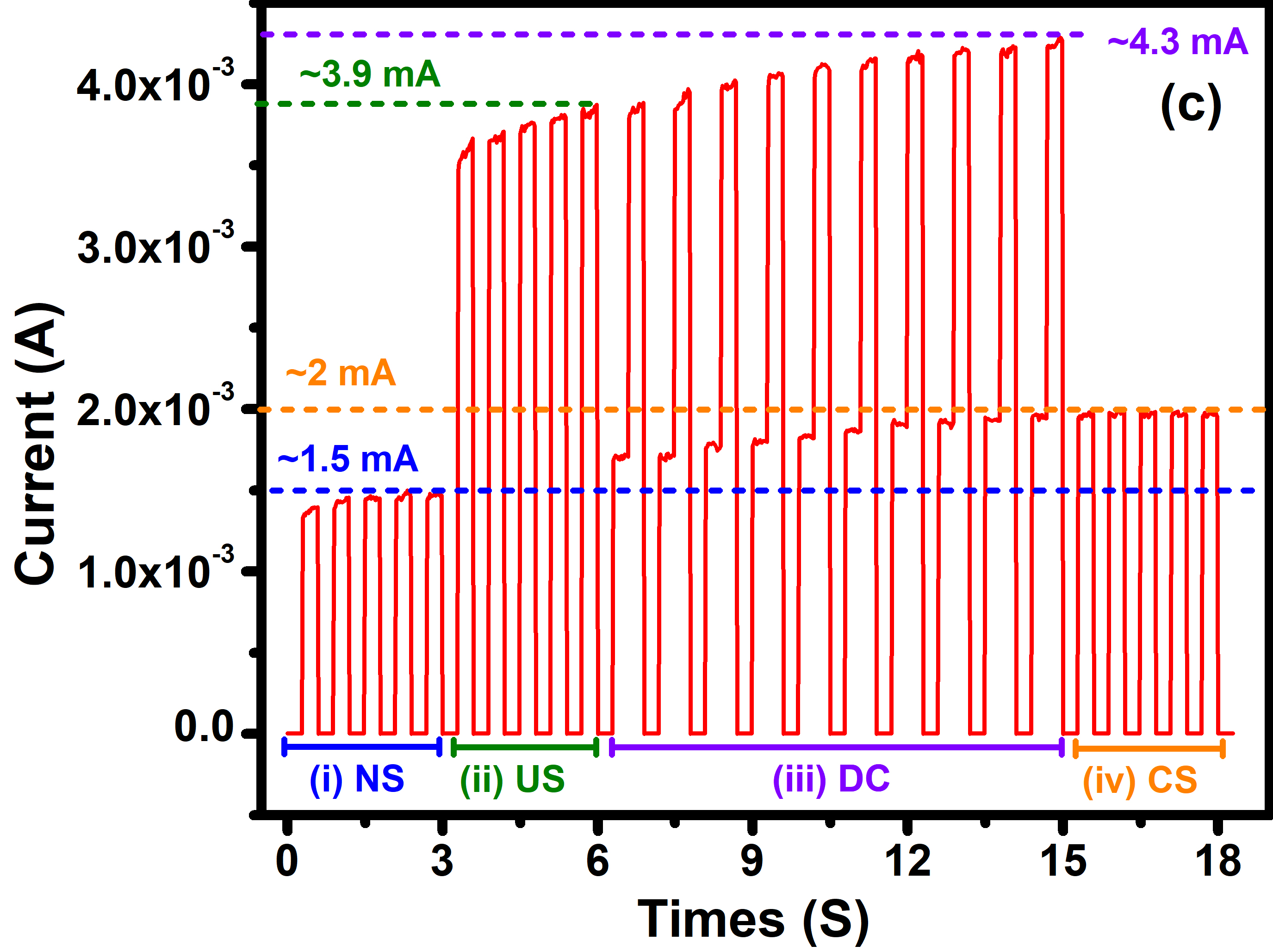}

 \caption{Experimental demonstration of Pavlovian associative learning using Ag/TiO$_x$/NiO$_x$/P$^{++}$-Si artificial synapse-- (a) A schematic illustration of the classical conditioning process depicting the initial neutral stimulus (NS), the unconditioned stimulus (US), the conditioning stage (DC) and the conditioned stimulus (CS); (b) Electrical pulse sequence used to emulate the food and ringing bell stimuli where the initial +1.5 V pulses represent NS, +2 V pulses represent US, the sequential combination of +1.5 V and +2 V pulses corresponds to the conditioning stage (DC) and the final +1.5 V pulses represent CS; (c) Output current responses of the device recorded during the NS, US, DC and CS stages, respectively demonstrating successful associative learning.}
  \label{Pavlov}
\end{figure*}

\begin{figure}[h]
    \centering
    \includegraphics[width=0.5\textwidth]{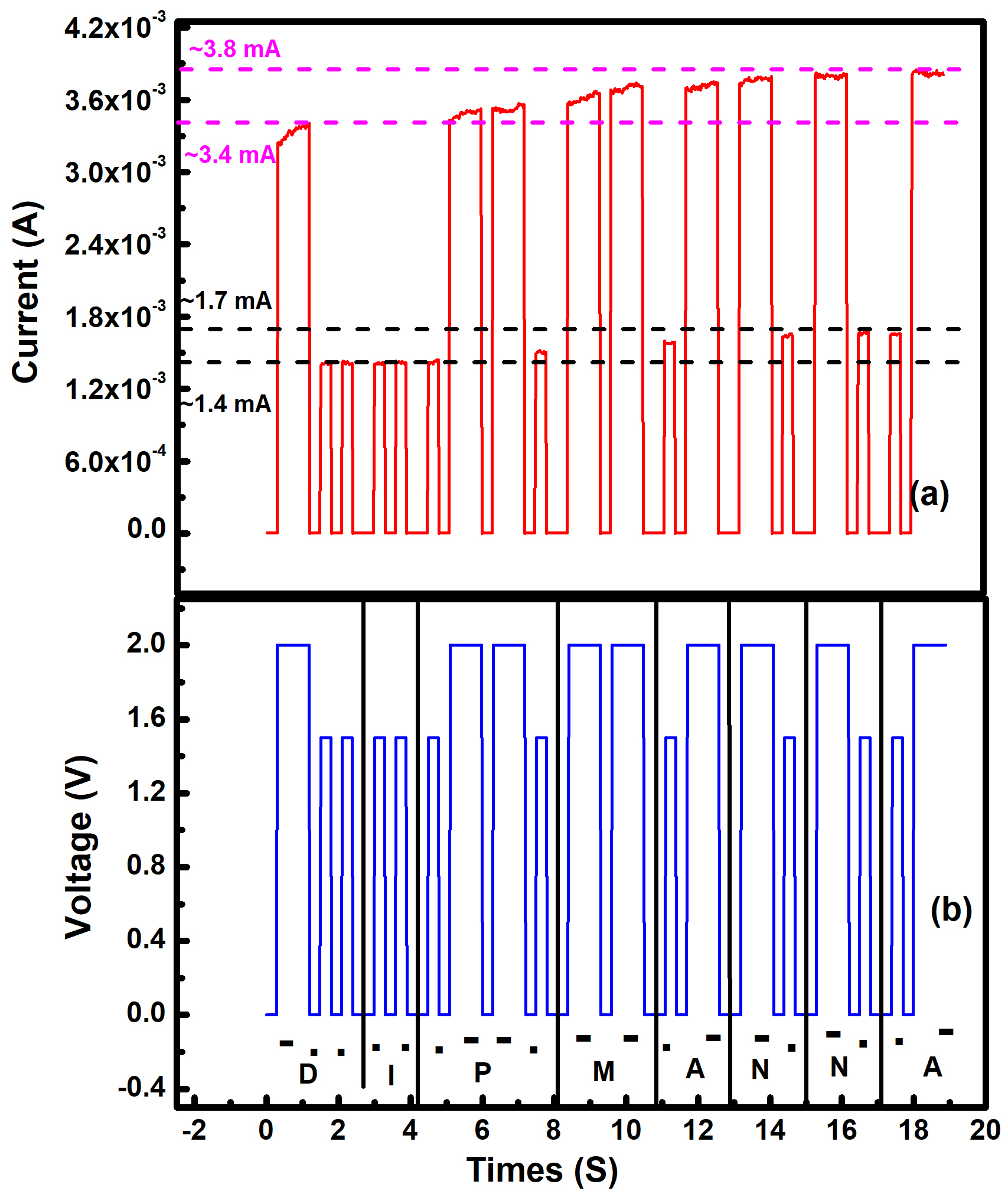}
    \caption{Experimental demonstration of Morse code encoding for the word "DIPMANNA'' using electrical pulses, where short and long voltage pulses represent dots ($\cdot$) and dashes ($-$) respectively.}
    \label{morse}
\end{figure}

\section{Application}\label{sec3}
\begin{figure*}[t]
    \centering
    \includegraphics[width=0.9\textwidth]{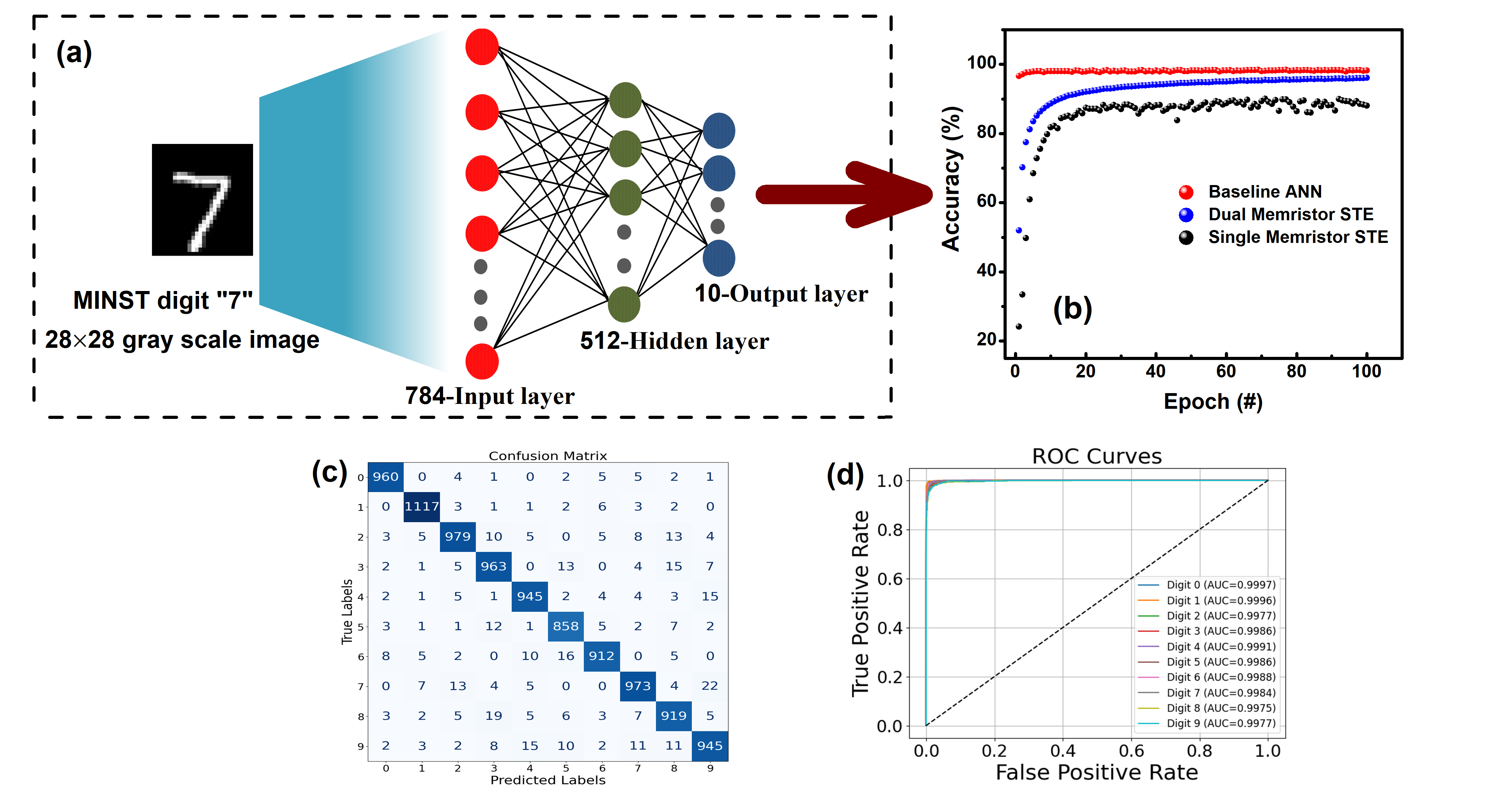}
    \caption{(a) A schematic diagram of the MNIST handwritten digit recognition process using the proposed neuromorphic system; (b) Evolution of training accuracy over 100 epochs with 3-point averaging except the Baseline  ANN; (c) Confusion matrix and (d) Receiver operating characteristic (ROC) curve.}
    \label{MNIST}
\end{figure*}

\subsection{Pavlovian Associative Learning}

Classical conditioning is one of the simplest forms of associative learning and demonstrates the ability of the brain to learn by forming associations between different stimuli. In the famous Pavlov’s dog experiment, food acts as the unconditioned stimulus (US) because it naturally causes salivation whereas, the sound of a bell is initially a neutral stimulus (NS) that does not produce any response. During conditioning (DC), the bell and food are presented together several times. As a result, the dog learns to associate the bell with the food. After conditioning, the bell alone becomes a conditioned stimulus (CS) and is able to trigger salivation which is known as the conditioned response (CR). This simple learning process has inspired the development of neuromorphic devices that can mimic associative learning in biological systems. The proposed memristor device was used to reproduce Pavlov's classical conditioning process by mimcking the associative learning behaviour of a biological dog's brain \cite{Jena, Minsu}. A schematic illustration of the classical conditioning process is shown in Figure \ref{Pavlov}(a). Figure \ref{Pavlov}(b) presents the sequence of five electrical pulses applied to the proposed heterostructure memristor to emulate associative learning. To perform the associative learning experiment, voltage pulses with a fixed pulse width of 300 ms and an interval of 300 ms were applied to the memristor device. A pulse of 1.5 V was used as the neutral stimulus (NS) whereas a pulse of  2.0 V was used as the unconditioned stimulus (US). The learning process was repeated for five consecutive cycles and the corresponding reponse current is shown in Figure \ref{Pavlov}(c). When the NS and US pulses were applied individually to the top electrode, the device shows threshold currents of approximately 1.5 and 3.9 mA, respectively. During the learning stage, each 1.5 V pulse was followed by a 2.0 V pulse with both pulses having a width of 300 ms. This pulse sequence represents the association between the neutral and unconditioned stimuli. As the paired pulses were repeatedly applied, the memristor gradually developed an associative learning response. Similar pulse based learning behaviour has also been reported in recent studies \cite{Jana}. Following five consecutive training cycles with the paired stimuli, the device reaches a maximum current of approximately 4.3 mA  which is higher than the threshold current obtained from the individual stimulus pulses. After the training process, only the neutral stimulus (+1.5 V) was applied using five consecutive pulses with a pulse width of 300 ms. In this case, the device produced a current of about 2 mA which exceeds the threshold current measured before training. This result indicates that the neutral stimulus alone was able to trigger the learned response, similar to the salivation response observed in Pavlov's classical conditioning experiment. To further verify the repeatability of this behaviour, the same experiment was performed using 10 consecutive pulses with a pulse width of 300 ms and a pulse interval of 300 ms  as shown in Figure S7. These results demonstrate that the Ag/TiO$_x$/NiO$_x$/p$^{++}$-Si memristor can successfully emulate Pavlov's classical conditioning behaviour.

\subsection{Morse Code Recognition}
To further demonstrate the capability of the proposed memristor for information processing, its performance on Morse code recognition was investigated. Morse code is a widely used communication system that represents letters and numbers using two basic signal patterns namely, a short pulse (dot) and a long pulse (dash) \cite{Shi}. By applying these pulse sequences to the memristor, the device was able to distinguish different Morse code patterns thereby demonstrating its potential for signal recognition and neuromorphic communication applications. Specific voltage pulses of 1.5 V and 2.0 V were assigned to represent the dot (.) and dash(-) signals, respectively. The dot signal was applied for 300 ms while the dash signal was applied for 900 ms. A zero bias interval of 600 ms was inserted after each pulse to indicate the space between Morse code symbols. These pulse sequences were used to encode and recognize different Morse code patterns using the memristor device. Based on the standard Morse code chart, the letters ``D'', ``I'', ``P'', ``M'', ``A'', ``N'', ``N'', and ``A'' were translated into their corresponding dot and dash patterns. The voltage pulse sequences used for each letter are presented in Figure \ref{morse}(b) and the measured current responses are shown in Figure \ref{morse}(a). The dot pulses produce current values in the range of 1.4--1.7 mA whereas the dash pulses generate higher current values of 3.4--3.8 mA. This clear difference in the current response allows the two signal types to be distinguishable without ambiguity. As the dot and dash pulses were applied in the correct sequence, the memristor generates distinct current patterns corresponding to each letter. Consequently, the complete word ``DIPMANNA'' was successfully encoded and recognized demonstrating the capability of the proposed memristor for information encoding and neuromorphic signal recognition. The reproducibility of the Morse code recognition capability was further examined using the word ``SCIENCE'' with the Ag/TiO$_x$/NiO$_x$/p$^{++}$-Si memristor. The corresponding voltage pulse pattern and the measured current response are presented in Figure S8. The successful recognition of this additional word confirms that the device can reliably decode different Morse code sequences with good repeatability.

\subsection{MNIST Handwritten Digit Recognition Using an Artificial Neural Network}

Evaluation of the neuromorphic computing capability of the fabricated Ag/TiO$_x$/NiO$_x$/p$^{++}$-Si synaptic device, handwritten digit recognition was performed using the Modified National Institute of Standards and Technology (MNIST) dataset. The experimentally measured non-linear conductance of the  memristor (LTP or LTD) has been used to update the synaptic weights of an ANN in response to the applied input patterns of hand written digits taken from MNIST database. The ANN architecture consisting of an input layer with 784 neurons, one hidden layer with 512 neurons and an output layer comprising of 10 neurons acts as a zero to nine digit classifier as shown in Figure \ref{MNIST}(a). In a memristor based ANN, synaptic weights are determined by the experimentally measured conductance pulse relationship. Since the measured conductance ($G_{meas}$) values are available only at integer numbers of the programming pulse count ($N_p$), a discrete forward and smooth gradient Straight Through Esimation (STE) is preferred for updating ANN weights. Following the algorithm, a differentiable surrogate function $z$ $\widetilde{G}(\cdot)$ with respect to a latent continuous variable is constructed from the measured conductance data. The synaptic weight may, therefore, be expressed as,
\begin{equation}
    w_{ij} = \widetilde{G}(z_{ij})
\end{equation}
where $z_{ij}$ denotes the continuous latent pulse variable associated with the
synaptic weight connecting neuron $i$ to neuron $j$ and may be expressed as,
\begin{equation}
  z_{ij} \in [0,N_{\max}]
\end{equation}
where $N_{\max}$ is the maximum number of programming pulses. The LTP model function is used here to represent the differentiable surrogate function $\widetilde{G}(\cdot)$ as \cite{Zhu},
\begin{equation}
\widetilde{G}(z_{ij}) = G_0 + A \left(1 - e^{-\frac{z_{ij}}{\tau}}\right)
\end{equation}
where LTP Nonlinearity $\tau$ = 12.369013042122612, $G_0 = G_{min}$ and 
$A = \left(G_{max} - G_{min}\right)/ \left(1 - e^{-\frac{N_{max}}{\tau}}\right)$.  
$G_{max}$ and $G_{min}$ are estimated from $G_{meas}$. During forward pass, weights are picked up from measured conductance look-up table after mapping the latent pulse variable to integer programming pulse count: \(N_{p,ij} = [round\left(z_{ij}\right)]_0^{N_{max}}\)
While in the backpropagation, smooth physics informed gradient \(\frac{d\widetilde{G}}{dz}= \frac{A}{\tau}e^{-z/\tau} \) with respect to the latent pulse variable $z$ is computed for updating the respective weights. The loss  gradient ($L$) with respect to the latent pulse variable is obtained using the chain rule where the latent pulse variable is updated using gradient descent $z_{ij}^{(t+1)}$ as,
\begin{equation}
    z_{ij}^{(t+1)}=z_{ij}^{(t)}-\eta \frac{\partial L}{\partial w_{ij}}\frac{d\widetilde{G}(z_{ij}^{(t)})}{dz_{ij}}
\end{equation}
where $t$ denotes the training iteration. This single memristor-based ANN was trained on 60,000-digit images and tested on 10,000 un­seen samples to evaluate pattern recognition performance. As presented in Fig. \ref{MNIST}(b) (black), the recognition accuracy of approximately 90\% is achieved. A conventional ANN can accomodate both the positive and negative weights during its training phase. However, as the conductance of a memristor is alsways a positive quantity, a single memristor-based ANN is constrained to assume positive weights only thus limiting the accuracy of pattern recognition. The use of double memristors easily facilitates  negative weights by differential encoding. The forward-pass weight is calculated by using the expression,

\begin{equation}
w_{ij} = \alpha\left[ G_{\mathrm{meas}}\left(\left[round(z_{ij}^+)\right]_0^{N_{max}}\right) - G_{\mathrm{meas}}\left(\left[round(z_{ij}^-)\right]_0^{N_{max}}\right)\right]
\end{equation}
and the smooth physics informed gradient with respect to the latent pulse variable $z \in \left\lbrace z_{ij}^+, z_{ij}^-\right\rbrace$ is used for updating the respective weights during backpropagation. Where,
\begin{equation}
\frac{d\widetilde{G}}{dz_{ij}^+}= \alpha\frac{\partial L}{\partial w_{ij}}\frac{A}{\tau}e^{-z_{ij}^+/\tau} 
\end{equation}
and
 \begin{equation}
\frac{d\widetilde{G}}{dz_{ij}^-}= -\alpha\frac{\partial L}{\partial w_{ij}}\frac{A}{\tau}e^{-z_{ij}^-/\tau}
\end{equation}
The differential encoding is found to improve accuracy to 95\% as shown in Figure \ref{MNIST}(b) (Blue) which is close to the ideal performance of 98\% achieved by the standard 784-512-10 classification network as presented by the red curve in Figure \ref{MNIST}(b) . The confusion matrix in Figure \ref{MNIST}(c) shows very few misclassifications for each digit which is further substantiated by the corresponding individual ROC curves as depicted in Figure \ref{MNIST}(d). The area under the curve (AUC) for each digit appears to account for more than 99\% of the total area. The demonstrated recognition performance verifies the capability of the fabricated memristive synapse so as to implement hardware oriented neural networks indicating its suitability for neuromorphic pattern recognition systems. The key performance parameters of the fabricated device are compared with those of recently reported oxide-based bilayer memristors and summarized in Supporting Information as Table--1. This comparison highlights the competitive switching and neuromorphic characteristics of the proposed Ag/TiO$_x$/NiO$_x$/p$^{++}$-Si heterostructure.

\section{Conclusion}
\label{sec3}
In summary, a non-volatile Ag/TiO$_x$/NiO$_x$/p$^{++}$-Si memristor was successfully fabricated using DC reactive magnetron sputtering followed by thermal evaporation of Ag as top electrodes. The fabricated device exhibits stable bipolar resistive switching with excellent endurance and retention characteristics. Structural, morphological and chemical analyses confirm the successful formation of the heterostructure whereas post switching TEM observations provide the evidence of Ag migration corroborating the proposed filament assisted switching mechanism in conjunction with the $p-n$ heterojunction. The device exhibits gradual conductance modulation during both potentiation and depression processes making it suitable for neuromorphic computing applications. Further, it successfully demonstrates short-term memory (STM) to long-term memory (LTM) transition through enhanced excitatory postsynaptic current (EPSC) under repetitive stimulation closely resembling biological synaptic learning. The memristor further emulates higher order neuromorphic functionalities that include Pavlovian associative learning, Morse code encoding and decoding and artificial nociceptor behaviour. Furthermore, the experimentally obtained long term potentiation (LTP) and long term depression (LTD) characteristics were employed for MNIST handwritten digit recognition and  a classification accuracy of ~95\% is achieved. These results highlight the strong potential of the proposed TiO$_x$/NiO$_x$  heterostructure as active lyaers of the memristor device as an efficient multifunctional platform for next generation non volatile memory and neuromorphic computing applications.
\section{Experimental details}
\label{sec4}
Ag/TiO$_x$/NiO$_x$/p$^{++}$-Si memristor devices were fabricated on highly doped p$^{++}$-Si wafer having resistivity of 0.0005 -- 0.005 $\Omega\cdot$cm  diced into 1 cm $\times$ 1 cm pieces after due cleaning using the standard RCA cleaning procedure. To remove the native SiO$_2$ layer, the cleaned substrates were immersed in a 1\% HF solution for one min and then rinsed in DI water and blow dried with high purity N$_2$ gas and immediately transferred into the sputtering chamber to minimize surface reoxidation. The thin films were deposited using a dual gun DC reactive magnetron sputtering system. 99.999\% pure Ni and Ti metal targets were mounted on two DC sputtering guns. After attaining a base pressure of $6 \times 10^{-6}$ mbar, Ar and O$_2$ gases were introduced into the chamber to generate the reactive plasma. The NiO$_x$ bottom layer was first deposited at room temperature using a DC power of 150 W with an Ar:O$_2$ flow ratio of 5:1. The chamber pressure was maintained at 5 mTorr during deposition and the deposition time was fixed at 15 min. After completion of NiO$_x$ deposition, the plasma of Ti target was ignited and TiO$_x$ layer was deposited on the already deposited NiO$_x$ film. The TiO$_x$ deposition was carried out mainating similar conditions. During both deposition processes, the substrate was continuously rotated at 30 rpm to ensure film uniformity. Metal Ag of $\sim$ 300 nm thick deposited  using thermal evaporation system on TiO$_x$/NiO$_x$/p$^{++}$-Si and patterned to get circular top electrodes to get Ag/TiO$_x$/NiO$_x$/p$^{++}$-Si device architecture. The DC current--voltage properties of the fabricated devices were masured using Keithley 2450 source meter with python-based control program.  Pulse induced switching measurements on the devices were carried out with a Keithley 2636 source meter where Keithley make KickStart software generates pulses of required widths and voltages. The structural properties of the heterostructure were examined by x-ray diffraction (XRD) using an Rigaku diffractometer over a 2$\theta$ range of 10$^\circ$--65$^\circ$. Surface morphology was characterized using a field emission scanning electron microscope (FESEM, Supra 40, Carl Zeiss, Germany) and an atomic force microscope (CSI, Nano-Observer). The chemical composition and electronic states were investigated by x-ray photoelectron spectroscopy (XPS, VSW Ltd., UK). The XPS measurements were carried out at a base pressure of $\leq 5 \times 10^{-9}$~mbar with Al K$_\alpha$ x-ray source ($h\nu = 1486.6$ eV) and the emitted photoelectrons were analyzed using a 150 mm hemispherical electron energy analyzer and pass energy of 20 eV. Cross-sectional microstructural characterization was performed using a transmission electron microscope (FEI Tecnai G2 F30-ST, USA) operated at a voltage of 300 kV. The microscope was equipped with a high-angle annular dark-field (HAADF) detector and an energy-dispersive x-ray spectroscopy (EDS) system enabling both high-resolution imaging and elemental mapping of the fabricated heterostructure.



\section{Acknowledgments}
Authors wish to acknowledge Professor Mrinmay Mukhopadhyay for providing x-ray  facility. The authors also acknowledge Mr. Debraj Dey, Mr Souvik Jana for XPS and AFM measurements, respectively. The first author wishes to acknowledge University Grants Commission for providing the research fellowship. 

{}

\end{document}